# Geo-FuB: A Method for Constructing an Operator-Function Knowledge Base for Geospatial Code Generation Tasks Using Large Language Models

Shuyang Hou[a], Anqi Zhao[a] ,Jianyuan Liang[a], Zhangxiao Shen[a], Huayi Wu[a*]

State Key Laboratory of Information Engineering in Surveying, Mapping and Remote Sensing, Wuhan University, Wuhan 430079, China

*Corresponding author: Huayi Wu, wuhuayi@whu.edu.cn

**Abstract:** The rapid emergence of spatiotemporal data and the growing need for geospatial modeling have led to the tendency of automating these tasks using large language models (LLMs), for research efficiency and productivity. The general LLMs encounter coding hallucinations in geospatial code generation due to the lack of domain-specific knowledge about geopsatial functions and relative operators. The retrieval-augmented generation (RAG) technique, coupled with an external operator-function knowledge base to provide description for these geospatial functions and operators, presents an effective solution to this challenge. However, a broadly recognized framework to develop such knowledge base is still absent. This study proposes a comprehensive framework to build the operator-function knowledge base, leveraging the rich semantic and structural knowledge embedded within geospatial scripts. The framework contains three core components: Function Semantic Framework Construction (Geo-FuSE), Frequent Operator Combination Statistics (Geo-FuST), and Combination and Semantic Framework Mapping (Geo-FuM). Geo-FuSE integrates techniques including Chain-of-Thought (CoT), TF-IDF, t-SNE, and Gaussian Mixture Model to uncover semantic features in the scripts. Geo-FuST employs abstract syntax trees (ASTs) and the APRIORI algorithm to identify frequent operator combinations from the code structures. Geo-FuM combines LLMs with the fuzzy matching algorithm to align these operator combinations with the function semantic framework, and forms the Geo-FuB knowledge base. An instance of Geo-FuB has been developed based on 154,075 Google Earth Engine scripts and available on https://github.com/whuhsy/Geo-FuB. A comprehensive evaluation metric is introduced to assess both the structural integrity and semantic accuracy of the framework. The overall accuracy reaches 88.89%, with structural accuracy reaching 92.03% and semantic accuracy at 86.79%. In addition, this study highlights the potential of Geo-FuB for optimizing geospatial code generation tasks based on RAG paradigm and fine-tuning paradigm, respectively, offering both research inspiration and empirical resources to address the limitations in such tasks.

**Keywords:** Code Generation; Large Language Models; Prompt Engineering; Retrieval-Augmented Generation; Google Earth Engine; Knowledge Base

## 1. Introduction

With the growing need for comprehensive and detailed understanding of geospatial processes, spatiotemporal analysis has become increasely complex. Geospatial analysis platforms—such as Google Earth Engine (GEE), ArcGIS, QGIS, and ENVI—have become essential tools to build geospatial models and perform spatiotemporal analysis[1]. Confronted with vast amounts of spatiotemporal data and the increasing complexity of modeling requirements, researchers are seeking to escape from the tedious and repetitive task of crafting basic codes. Automatic geospatial code generation has emerged as a necessary strategy to simplify and streamline the geospatial modeling process[2].

Geospatial code generation is a knowledge-intensive task, facing challenges such as ambiguous and non-standard user requirements, the complexity of interdisciplinary knowledge, and the data heterogeneity[3]. Large language models (LLMs), with their powerful contextual learning, logical reasoning, and natural language generation capabilities, offer the potential for automatic geospatial code generation[4]. However, LLMs are typically trained on extensive general-purpose corpora, which often fall in short of incorporating domain-specific expertise[5]. Moreover, due to the probabilistic nature of the generation in LLMs, their prediction mechanisms are characterized by dynamic variability, instability, inherent randomness, and weak controllability. While these biases have limited impact on natural language tasks, they can lead to significant errors in code generation, often known as "coding hallucinations[6]", such as confusion in coding language, the invention of non-existent operator functions, and the fabrication of operator names. As a result, there is considerable demand for enhancing the reliability of LLMs to generate geospatial code automatically.

An effective method for integrating domain-specific and task-specific insights into LLMs is to integrate the Retrieval-Augmented Generation (RAG) technology with customized external knowledge base to guide the LLM in generating more accurate results[7]. However, creating an external knowledge base for geospatial modeling tasks that covers comprehensive operator combinations, function descriptions, and domain-specific logic requires not only precise and complete content but also well-organized and easily retrievable structure[8]. Currently, a widely-recognized framework or protocol that can guide the development and assessment of these specialized knowledge bases is still lack. Generally, user-developed geospatial scripts in different geospatial analysis platforms include two principal categories of knowledge that are crucial for geospatial modeling[9]: (1) **Semantic Information,** including function descriptions and process annotations, which reveal the purpose and operational logic of the scripts**. (2) Structural Knowledge,** encompassing computational logic, dependencies between operators, and combination patterns[10]. However, to the best of our knowledge, these geospatial scripts have not yet been systematically mined, organized, and utilized for constructing domain-specific knowledge bases in the geospatial modeling field.

To address the issues faced by LLMs in geospatial code generation, including the coding hallucination, the lack of specialized knowledge bases, and insufficient mining of geospatial script knowledge, this paper proposes a framework named Geo-FuB to construct an operator-function knowledge base for geospatial code generation tasks using LLMs. This framework includes three core components: **Function Semantic Framework Construction (Geo-FuSE)**, **Frequent Operator Combination Statistics (Geo-FuST),** and **Combination and Semantic Framework Mapping (Geo-FuM)**. **Geo-FuSE** integrates the Chain of Thought (CoT), TF-IDF, t-SNE, Gaussian Mixture Model (GMM) and expert experience to uncover semantic features in different geospatial scripts. **Geo-FuST** employs abstract syntax trees (ASTs) and the APRIORI algorithm to identify frequent operator combinations from the code. **Geo-FuM** leverages LLMs and the fuzzy matching algorithm to align various operator combinations with the proposed semantic framework, and utimately form the operator-function knowledge base. A instance of the Geo-FuB knowledge base has been constructed based on 154,075 geospatial scripts in GEE environment and is available on the https://github.com/whuhsy/Geo-FuB. Furthermore, a comprehensive evaluation metric was introduced to assess both the structural integrity and semantic accuracy of the developed knowledge base, which shown an overall accuracy of 88.89%, with structural accuracy reaching 92.03% and semantic accuracy at 86.79%.

The contributions of this study are summarized as follows:

- The Geo-FuB framework is proposed for constructing an operator-function knowledge base for geospatial code generation tasks using LLMs. This framework contains three core methods: Geo-FuSE, Geo-FuST, and Geo-FuM, which is applicable for constructing such operator-function knowledge bases across various geospatial analysis platforms.

- Based on 154,075 geospatial scripts from the GEE platform, this paper developed an instance of the Geo-FuB, including a semantic framework with 3 primary categories, 8 secondary categories, and 21 tertiary categories, 4350 frequent operator combinations, and their complete knowledge base mapping results. This knowledge base is available on https://github.com/whuhsy/Geo-FuB for usages in code generation and other downstream tasks.

- A comprehensive evaluation metric is proposed considering both structural consistency and semantic correctness. The evaluation results show an overall accuracy rate of 88.89% (structural accuracy of 92.03% and semantic accuracy of 86.79%). This evaluation metric can also be used to evaluate the knowledge base constructed for geospatial analysis platforms based on the Geo-FuB.

- Through the discussion of specific application scenarios, this study showcases the potential of Geo-FuB in optimizing geospatial code generation tasks within RAG paradigm and fine-tuning research paradigms, thus providing research inspiration for domain scholars.

The organization of this paper is as follows. Section 2 reviews the specialized applications of LLMs, the application of RAG technology in geospatial domain, and related research on the construction of domain-specific knowledge bases for geospatial tasks. Section 3 provides a detailed design and implementation of the Geo-FuB. Section 4 presents the instance of the Geo-Fub knowledge base using GEE scripts and conducts quantitative evaluation. Section 5 explores the potential value of Geo-FuB for LLMs through specific application scenarios and discusses its limitations. The summarization and the application prospects of this paper are provided in Section 6.

## 2. Related Work

### 2.1. Domain Specialization of Large Language Models

Pre-trained language models (PLMs) based on self-attention mechanism and Transformers can extract general language representations from large-scale unlabeled data in an unsupervised manner[11]. These models have demonstrated significant advantages across a spectrum of natural language processing (NLP) tasks[12], such as text generation[13,14], machine translation[15], question answering system[16-18], sentiment analysis[19], and knowledge extraction[20]. As the scale of model parameters has grown from millions to billions, the contextual understanding and logical reasoning capabilities of PLMs have significantly improved[21]. These large-scale PLMs are referred to as large language models (LLMs).

LLMs overwhelm other small-scale PLMs owing to the extensive corpora used in their training process. However, the large-scale parameters and extensive training data act as a "double-edged sword" for LLMs[22]. Due to the substantial computational resources required for pretraining, LLMs cannot acquire the latest information in real time, which leads to serious knowledge gaps. Additionally, since general corpora are larger than domain-specific data, when faced with highly-specialized tasks, the models may produce seemingly correct but misleading answers, known as "knowledge hallucination[23]". Therefore, integrating domain knowledge to LLMs and promoting their specialization in domain[24](i.e., the "LLM+X" model) has become a research focus.

Based on domain-specific adjustments, LLMs have been applied to specialize tasks of various fields, including social sciences (such as education[25], finance[26] and law[27]), natural sciences (such as biomedicine[28] and earth sciences[29]), and applied sciences (such as human-computer interaction[30], software engineering[31], and cybersecurity[32]). Examples of these applications include: (1) **Advanced Information Extraction, aiming at** identifying entities, relationships, and events from domain-specific texts, such as recognizing genes from biomedical literature[33] or detecting legal clauses in contracts[34]; (2) **Text Generation and Summarization**, which **produces** high-quality, domain-specific content and creates accurate summaries for complex domain-specific texts; (3) **Data-Driven Prediction and Recommendation that analyzes** domain-specific data for making predictions and provides recommendations, such as predicting financial trends[35] or proposing personalized medical treatment plans[36]. **(4) Conversational Agents and Expert Systems,** integrating LLMs into conversational agents or expert systems to provide domain-specific guidance, such as virtual teachers[37] or legal chatbots[38].

### 2.2. Geospatial Code Generation with Large Language Models

The geospatial domain faces unique challenges in leveraging LLMs for automated geospatial modeling becasue of the highly specialized code requirements in different geospatial analysis platforms. Automatic geospatial modeling requires LLMs not only to handle complex geospatial data but also to generate code that conforms to platform specifications (such as GEE[39]). These platforms typically provide programming environments based on languages like Python or Java and have developed specifications and operational logics for geospatial analysis, which necessitates LLMs to take extra efforts to learn and adapt.

LLMs have been used in automatic code generation[41,42], such as generating or analyzing code based on natural language descriptions[43], identifying errors[44], or proposing improvements, nevertheless, existing research predominantly focused on general code generation and has not been specifically adapted to the unique code of geospatial platforms. The foundational training corpora of LLMs mainly involve general programming knowledge and lack the specific function libraries and operational workflows unique in different geospatial analysis platforms. The knowledge gap makes it challenging for the LLMs to generate code that meets the requirements of these geospatial analysis platforms. Even though techniques like prompt engineering and CoT have improved the accuracy and flexibility of LLMs, their effectiveness remains inadequate when faced with the specialized needs of the geospatial analysis.

### 2.3. Construction of Specialized Knowledge Bases for Geospatial Tasks

The generation and reasoning performance of LLMs depend on the quality of their training corpora[45]. It is crucial to construct target-oriented and high-quality knowledge bases to enhance the models' generative capabilities for the geospatial field. Researchers have integrated specialized knowledge bases to optimize the performance of LLMs in downstream tasks. For instance, Cheng Deng et al. compiled 6 million geoscience academic papers to create the GeoSignal dataset, which includes article content, classifications, references, and entity information, providing valuable resources for tasks such as geospatial question answering, named entity recognition and conceptual relationship reasoning[46]. Similarly, Yifan Zhang et al. developed the BB-GeoPT knowledge base, which combines high-quality GIS journal papers, Wikipedia GIS pages, general instruction sets, and specialized instruction data to enhance the performance of geospatial knowledge question-answering systems[47].

Although these efforts have advanced the development and application of geospatial knowledge bases, a comprehensive and systematic knowledge base for automated geospatial modeling has not yet been established. Such knowledge base should establish a complete and retrievable knowledge system that covers independent operator knowledge of geospatial analysis platforms, geoprocessing function semantics, operator combination information and function-operator mappings. Incorporating these corpora into the training or inference process of LLMs can improve the accuracy and efficiency of geospatial code generation, thereby enhancing the models' automatic modeling capabilities and providing more intelligent and efficient solutions for geospatial research.

Crowdsourced geospatial scripts are invaluable resource[48] that include both the subjective semantic insights provided by detailed function descriptions and process annotations, along with the objective statistical data that maps out the intricate dependencies and combination patterns of operators. These information can be transformed into retrievable knowledge and harnessed as vital input for LLMs in terms of retrieval augmentation, generation optimization, and fine-tuning processes.

## 3. Methodology

### 3.1. Overall Framework

This study proposes a comprehensive framework for constructing an operator-function knowledge base. The complete workflow of the knowledge base construction is illustrated in Figure 1, which encompasses the three main components: Geo-FuSE for Function Semantic Framework Construction (the green part), Geo-FuST for Frequent Operator Combination Mining (the blue part), and Geo-FuM for Combination and Semantic Framework Mapping (the teal part).

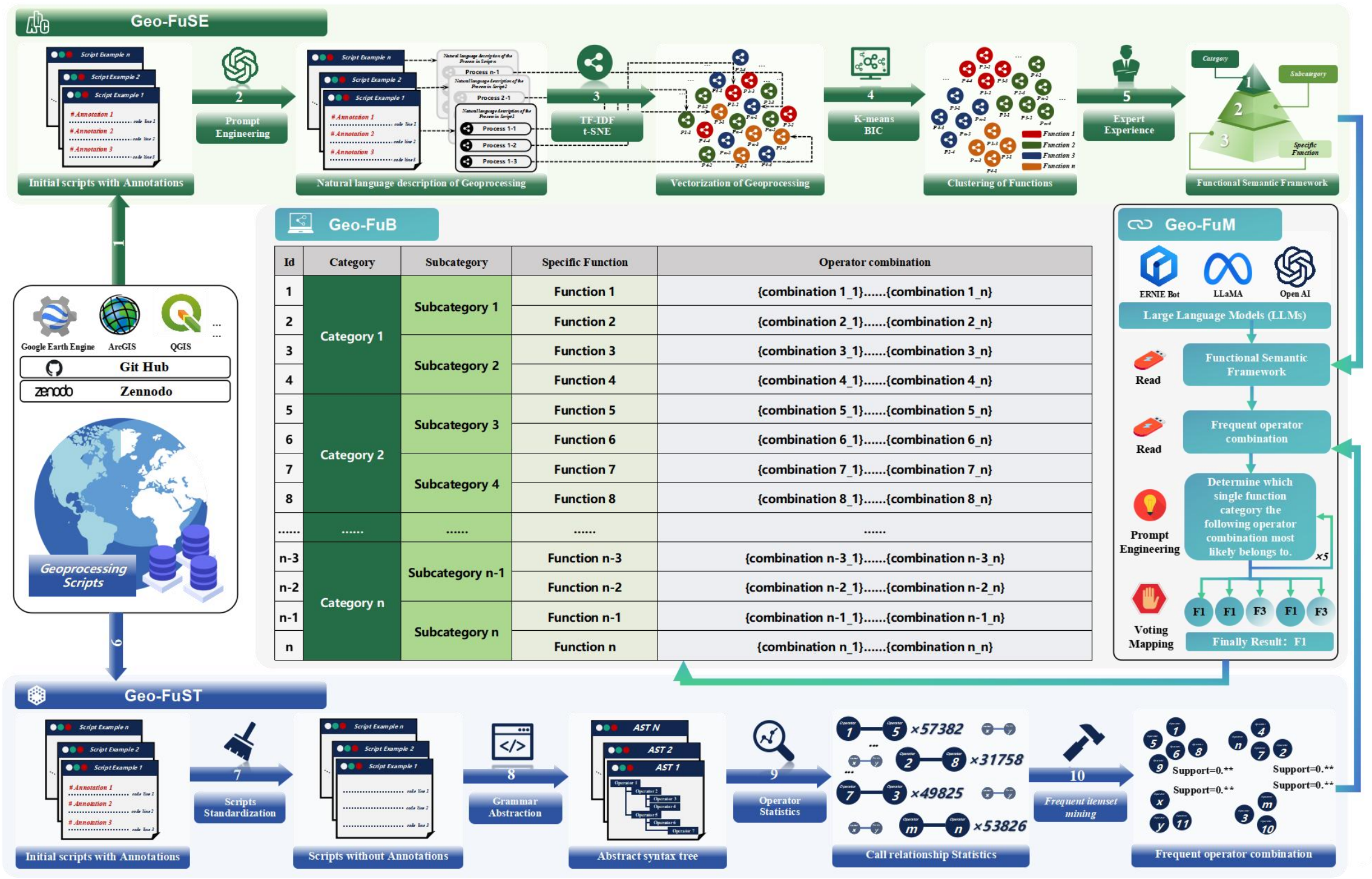

Fig.1 Workflow of the Geo-FuB Framework Construction

In the **Function Semantic Framework Construction** stage, geospatial scripts are treated as domain-specific natural language texts. Through prompt engineering with LLMs, the workflows within the scripts are distilled, while the raw code with annotations are parsed into a collection of functional statements (i.e., natural language descriptions of a series of geospatial operations within each script). Next, the TF-IDF algorithm is used to construct vector matrices for these functional statements. The vector matrices are reduced to two dimensions based on t-SNE, and clustered through the Gaussian Mixture Model (GMM) and Bayesian Index Criterion (BIC) index. The clustered results are then refined and semantically labeled with expert experience to construct a three-level standard Function Semantic Framework, Geo-FuSE.

In the **Frequent Operator Combination Mining** stage, geospatial scripts, after annotation cleaning, are transformed into JSON format using Abstract Syntax Tree (AST) technology. A traversal algorithm is used to calculate the call relationships and frequencies between operators, and generate an operator call frequency table. By setting a minimum support threshold, the Apriori algorithm is then employed to mine frequent operator combinations from this table, which forms the Frequent Operator Combination, Geo-FuST.

In the **Combination and Semantic Framework Mapping** stage, a prompt is designed and four advanced LLMs are selected for five consecutive inquiries. Using a voting mechanism through single-model or model combination schemes, the final mapping results are chosen, and ultimately completing the construction of the knowledge base.

### 3.2. Geo-FuSE: Function Semantic Framework Construction

In geospatial code generation tasks, a function is defined as a specific task or operation accomplished through the collaboration of a set of operators. A function in the geospatial script contains the following characteristics: **Clarity:** Each function is composed of a set of operators and has a clear boundary and objective. **Operability:** Functions can be refined into specific operational steps, and accomplished through the implementation of operators. **Hierarchical Structure:** Functions are defined in a layered manner, with primary, secondary, and tertiary categories, which facilitates systematic classification, indexing, and hierarchical evaluation. The overall workflow of Geo-FuSE is illustrated in Figure 2.

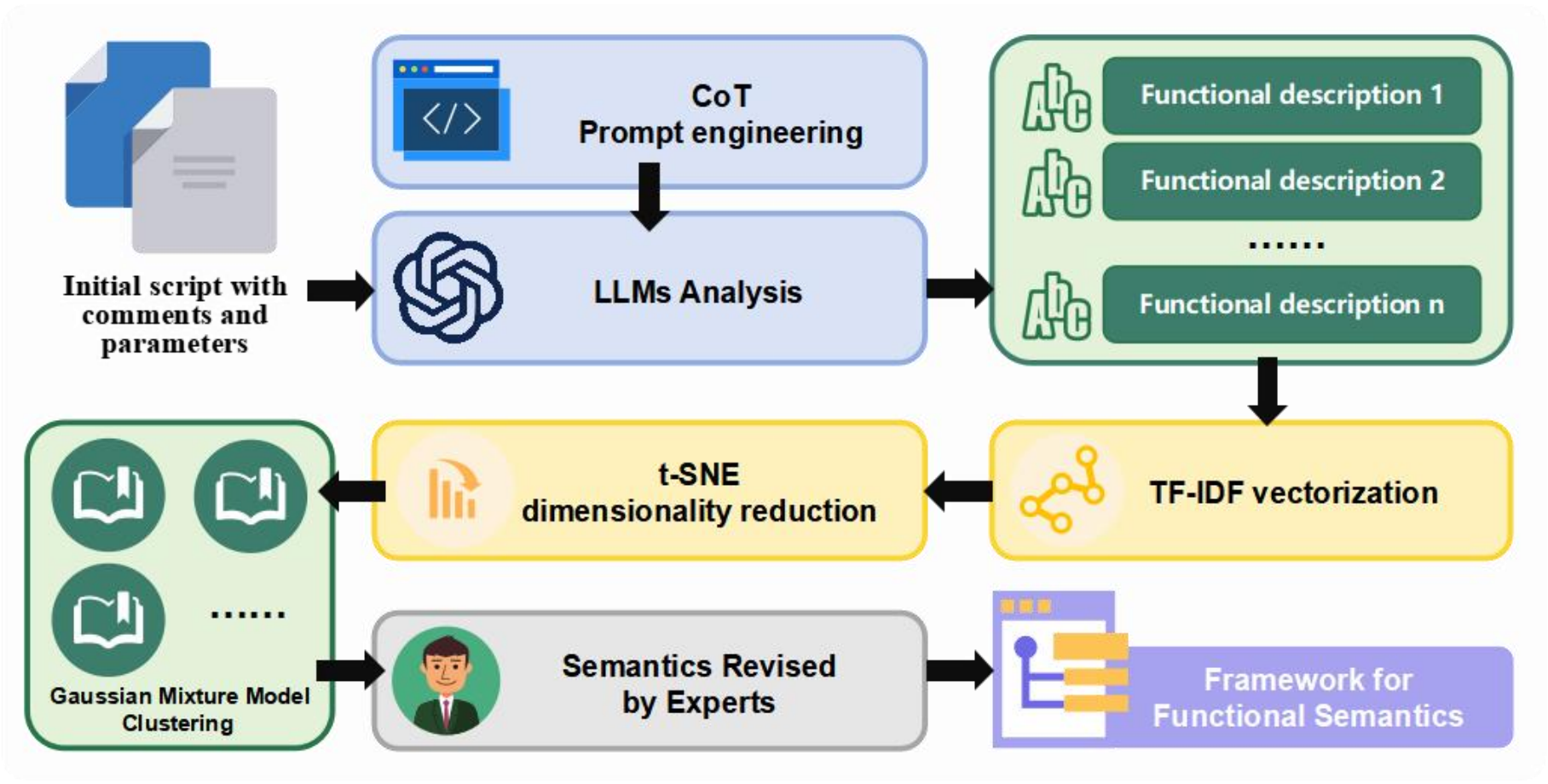

Fig.2 Workflow of the Geo-FuSE

### 3.2.1. CoT-based Prompt Engineering for Function Semantic Extraction

Geospatial scripts usually contain rich functional semantic[49] annotations for the design logic, code functionality explanation, dataset characteristics, and research objectives. Meanwhile, the operator names in the code can also implicitly convey functional semantics that are difficult to parse with traditional language models. Long scripts usually contain multiple functional units for complex, multi-step tasks such as large-scale terrain data analysis, while short scripts mainly focus on single data processing objective such as meteorological parameter extraction and usually contain less or even one functional unit. This difference makes it challenging to delineate functional granularity. Fortunately, the powerful natural language parsing and semantic understanding capabilities of LLMs allow them to address this challenge and focus on deeper functional semantic analysis.

The Chain of Thought (CoT) method[50] simulates the thought process of human teaching by setting a series of highly relevant prompt words, and provides a heuristic thinking mode for LLMs to progressively generate the correct answer. By combining the CoT with a designated template that contains clear and structured prompting instructions, LLMs are able to accurately parse the set of functional statements. The design of the CoT-based prompt template is shown in Figure 3, with the words adaptable to different geospatial processing platforms and languages.

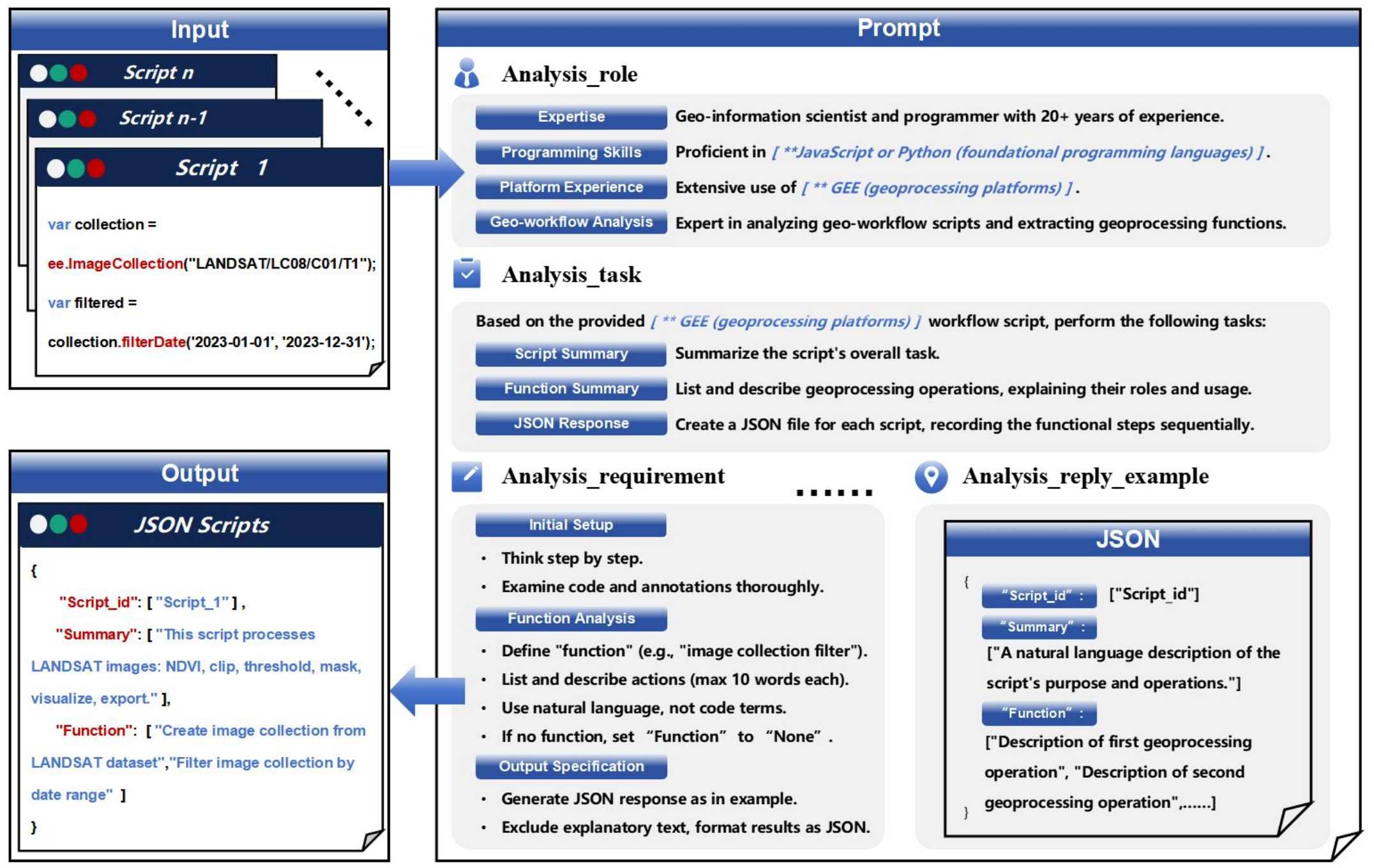

Fig.3 CoT-based Prompt Template

Specifically, the prompt template consists of four components. The **Analysis_role** defines the role and background knowledge of the language model as a geoinformation scientist and programmer with over 20 years of experience, proficient in JavaScript or Python and geospatial processing platforms such as Google Earth Engine (GEE). The **Analysis_task** clarifies the tasks that the LLM needs to complete, including summarizing the overall task of the script, listing and describing the geoprocessing and analyzing operations within the script, and generating a JSON file that records the functional steps. The **Analysis_reply_example** provides a standard reply example that illustrates the purpose and operations of the script and detailed descriptions for each function, which ensures that the content and format of the output exactly meets the requirements. Finally, the **Analysis_requirement** lists the detailed steps and requirements following the CoT thought, which guides the LLM to complete the tasks step-by-step.

### 3.2.2. TF-IDF Feature Matrix

The extraction of functional workflows primarily relies on the internal knowledge generation of LLMs, which poses challenges while producing semantically-correct descriptions because of the diversity in expressions. The first challenge is the issue of **semantic redundancy**, as the same function may be described significantly differently in various scripts, thus causing redundancy. For example, the "data retrieval" function might be described as "perform data retrieval operation," "conduct data search," or "retrieve and extract data.". The second challenge is the **ambiguity of functional boundaries**. For example, "data analysis and visualization" and "data visualization display" are similar in semantics, they can be considered either different aspects of the same function or divided into two separate functions. To address these issues, the natural language descriptions of functional statements are converted into vector representations in a high-dimensional space, thereby quantifying the subtle semantic differences between functional statements at a granular level.

TF-IDF (Term Frequency-Inverse Document Frequency) is a method to evaluate the importance of natural language segments within the entire collection of segments by calculating term frequency and inverse document frequency[51]. Specifically, TF (term frequency) measures the frequency of a particular term within a specific collection of natural language segments, while IDF (inverse document frequency) measures the rarity of a particular term across the entire collection of segments. By applying a logarithmic transformation, IDF reduces the weight of that appear frequently in multiple documents. The calculation formulas for both are as follows:

$$\mathrm{TF}(t,d)=\frac{f(t,d)}{N_d}\#(1)$$

$$\mathrm{IDF}(t)=\log\left(\frac{N}{1+n(t)}\right)\#(2)$$

Where $f(t,d)$ represents the frequency of the term t (i.e., a feature word in the parsed functional statements) in the document set d (i.e., the collection of all parsed functional statements), $\mathrm{N_d}$ represents the total number of terms in the document set d, N represents the total number of natural language statements in the document set, and $\mathrm{n(t)}$ represents the number of natural language statements containing the term t. The final TF-IDF value is obtained by multiplying TF and IDF, and through the normalization process of TF-IDF, all terms are represented in a standardized manner:

$$\text{TF-IDF}(t,d)=\mathrm{TF}(t,d)\times\mathrm{IDF}(t)\#(3)$$

$$v_{norm}=\frac{v}{||v||2}=\frac{v}{\sqrt{v_1^2+v_2^2+\cdots+v_n^2}}\#(4)$$

Through TF-IDF analysis, the importance and relevance of terms within the function descriptions are captured and quantified based on feature quantity and vector matrices. The vectorization of feature statements enhances the representation of the vocabulary, and provides data support for clustering and similarity calculation in subsequent stages.

### 3.2.3. t-SNE Dimensionality Reduction

Since the functional statements are constructed into high-dimensional vectors after the TF-IDF, directly clustering high-dimensional vectors will consume substantial computational resources and be difficult to be visualized. Therefore, the dimensionality reduction method is applied as a prevalent approach to understand the inherent connections between functional statements represented in high-dimensional vectors.

Among the dimensionality reduction methods, t-SNE (t-Distributed Stochastic Neighbor

Embedding) outperforms others for dissecting intricate datasets due to its proficiency in maintaining the local data structure of high-dimensional information[52]. As a nonlinear technique, t-SNE is especially suitable for the visualization of high-dimensional data. It embeds high-dimensional data into a low-dimensional space by maximizing the probability that similar data points are adjacent in the low-dimensional space, thereby visually presenting the data structure.

Specifically, t-SNE first constructs a probability distribution for each data point in the high-dimensional dataset, reflecting the similarity between data points. For data points $x_i$ and $x_j$ in the high-dimensional space, their similarity is represented by the conditional probability $p_{j|i}$:

$$p_{j|i} = \frac{exp\left(-\| x_i - x_j\|^2/2\sigma_i^2\right)}{\sum_{k\neq i} exp\left(-\| x_i - x_k\|^2/2\sigma_i^2\right)} \#(5)$$

Where $\sigma_i$ is the width of the distribution, controlling the range of neighboring points considered. To be symmetric, the algorithm defines:

$$p_{ij} = \frac{p_{j|i} + p_{i|j}}{2N} \#(6)$$

Where N is the total number of data points. In the low-dimensional space, t-SNE assigns a corresponding probability distribution $q_{ij}$ to each point, using a similar form:

$$q_{ij} = \frac{\left(1 + \| y_i - y_j\|^2\right)^{-1}}{\sum_{k\neq l} (1 + \| y_k - y_l\|^2)^{-1}} \#(7)$$

t-SNE optimizes the positions of data points in the low-dimensional space by minimizing the Kullback-Leibler (K-L) divergence between the high-dimensional and low-dimensional probability distributions:

$$C = \sum_{i\neq j} p_{ij} \log\frac{p_{ij}}{q_{ij}} \#(8)$$

The t-SNE process not only preserves the local similarities of the data but also emphasizes the separation between different clusters, making the aggregation and separation of functional terms more distinct in the visualization. By using t-SNE, potential functional groups and semantic boundaries can be identified, providing an intuitive reference for subsequent clustering analysis.

#### 3.2.4. Gaussian Mixture Model Clustering

The results from t-SNE dimensionality reduction are organized into groups with clear semantic boundaries through clustering analysis to classify similar term vectors. To achieve this, the Gaussian Mixture Model (GMM), a clustering method based on probabilistic statistical theory, is used for clustering analysis in this study[53]. GMM has significant advantages over other classic clustering methods, such as K-means and DBSCAN[54]. K-means requires a preset number of clusters and lacks an effective evaluation mechanism, making it difficult to determine the optimal value. Additionally, it is only suitable for clusters with relatively fixed sizes and densities, limiting its flexibility. DBSCAN can identify clusters of arbitrary shapes, but its performance is highly dependent on parameter settings, making it difficult to handle clusters with varying densities and high-dimensional data.{Fan, 2023 #6} GMM does not require a preset number of clusters and can automatically determine the best number of clusters within the desired range. Furthermore, GMM captures clusters of different shapes and sizes in the data set flexibly through the linear combination of Gaussian distributions.

The core idea of GMM is to consider the dataset as a complex distribution composed of multiple Gaussian distributions. Each Gaussian distribution in this context represents a potential cluster, each of which is a set of terms with similar functions. During the clustering process, GMM fits the data by estimating the parameters of each Gaussian distribution. **Mean vector** $\mu_k$ represents the center of the $k$-th cluster. **Covariance matrix** $\Sigma_k$, defines the shape and size of the $k$-th cluster. **Mixing weight** $\pi_k$, indicates the importance of the $k$-th cluster in the entire dataset, with $\sum_{k=1}^{K} \pi_k = 1$. Through the estimation of these parameters, GMM can flexibly adjust the linear combination of different Gaussian functions to accommodate the complexity and diversity of the data. The probability density function for a term vector $x_i$ is defined by:

$$p(x_i|\Theta) = \sum_{k=1}^{K} \pi_k \cdot \mathcal{N}(x_i|\mu_k, \Sigma_k)\#(9)$$

Where $\Theta = \{\pi_k, \mu_k, \Sigma_k\}$ represents the set of model parameters, $p(x_i|\Theta)$ represents the probability density of term vector $x_i$ given the parameter set $\Theta$, $K$ represents the total number of clusters, $\pi_k$ is the weight of the $k-th$ Gaussian distribution, and $\mathcal{N}(x|\mu_k, \Sigma_k)$ is the probability density function of a Gaussian distribution with mean $\mu_k$ and covariance matrix $\Sigma_k$. This formula

indicates that the generation probability of each term vector $x_i$ is determined by the weighted sum of multiple Gaussian distributions. Each term vector $x_i$ is influenced by contributions from multiple clusters, which provides a perspective of multi-cluster affiliation rather than a single one. By adjusting the weights $\pi_k$ , the influence of different clusters on $x_i$ can be precisely quantified. This reflects the overall explanatory power of the model for the term vector, representing the comprehensive fit of $x_i$ across all potential clusters. A high value of $p(x)$ indicates that the term vector $x_i$ is well-fitted, which indicates that the model can describe the characteristics of $x_i$ using all its Gaussian components, and highlights the soft clustering characteristic of GMM.

To find the best GMM parameters to describe the data, parameter estimation is continuously performed to maximize the likelihood function of the probability density. The log-likelihood function is used to simplify the calculation, where $N$ is the number of term vectors $x$:

$$\ln L(\Theta) = \sum_{i=1}^{N} \ln\left(\sum_{k=1}^{K} \pi_k \cdot \mathcal{N}(x_i|\mu_k, \Sigma_k)\right) \#(10)$$

For each specific value of K, the parameters of GMM are selected and refined to reach the model's optimal configuration for that K. Different K values reflect differing levels of model complexity and the data fitting effect. Identifying the optimal model among those defined by each K value is crucial for assessing performance. To ascertain the best-performing model under different K values and identify the most effective number of clusters, it is imperative to evaluate the optimal model for each K value. The number of clusters not only determines how term vectors are organized but also shapes the scope of the final functional semantic framework. The Bayesian Information Criterion (BIC) is used as the standard for model selection to identify the optimal number of clusters, which is a criterion that balances model fit and complexity:

$$\text{BIC} =- 2 \cdot \ln L(\Theta) + p \cdot \ln(N) \#(11)$$

Where $\ln L(\Theta)$ represents the maximum value of the log-likelihood function of the model, $p$ is the number of model parameters, reflecting its complexity, and $N$ is the number of samples. The number of clusters that minimizes the BIC value is selected as the optimal solution, as a lower BIC value usually indicates that the model achieves a good fit to the data while avoiding overfitting and excessive model complexity.

#### 3.2.5. Expert Experience-Based Classification

Expert experience is integrated to interpret semantic groups and establish a standardized function semantic framework based on clustering results. The framework adopts a three-level classification strategy based on comprehensive considerations of the following three dimensions:

**First, it follows the coding common sense and operational logic.** Geospatial data processing usually follows a workflow of “preprocessing-core operation-postprocessing”. Therefore, the framework sets this workflow as the first level to grasp the function categories at a macro level. The bottom level directly corresponds to the fine-grained functional units obtained from the clustering analysis, representing specific implementations. To connect these two levels, an intermediate second-level category is established to summarize and organize function units with similar properties or logical connections, making the framework more logical and complete.

**Second, the multi-level classification system supports precision and flexibility for future function mapping.** It constructs a multi-tiered matching mechanism that allows the system to automatically backtrack to the previous level when the finest-grained functional unit cannot be precisely matched, until a generalized match is achieved. This mechanism enhances the reliability and applicability of the mapping results.

**Finally, integrating the function semantic framework into the application scenario of the CoT methodology.** The framework not only provides a clear hierarchical structure but also supports a progressively deeper, step-by-step reasoning path. Each level of classification within the framework becomes a node in the chain, allowing for consecutive questioning that narrows down from broad categories to specific functional units. For instance, a user might inquire whether a certain function belongs to the first-level major category (e.g., preprocessing) at the begining, then further refine to the second level (e.g., data cleaning), and finally determine the specific third-level function (e.g., outlier removal). The CoT method matches human’s natural process of handling complex problems that progressively narrows down the scope to approach the answer, significantly enhancing the classification and decision-making accuracy and efficiency of LLMs.

### 3.3. Frequent Operator Combination Mining

A function is defined by a specific combination of operators, however, the operator combinations to achieve the same function can vary. For example, a function might be realized by the collaboration of operator A and operator B, or through the combination of operator C and operator D. Given the infinite possibilities of operator combinations, relying solely on manual enumeration has significant limitations. On one hand, theoretically feasible operator combinations might not meet practical application needs and thus be overlooked. On the other hand, effective combinations in practice might be missed due to the limitations of manual enumeration. Therefore, identifying and assessing operator combinations based on human experience is impractical and inefficient, and lacks the flexibility to find significant combinatorial patterns.

Therefore, a data-driven approach is adopted by analyzing a large number of scripts and using statistical methods to identify frequent operator combinations. The frequent occurrence of these combinations statistically indicates their importance and synergy in achieving specific functions and represents how users tend to use the operators in practical. This data-driven approach compensates for the shortcomings of the enumeration method and provides operator combination information for function framework construction, thus enhancing its value in LLM modeling tasks.

#### 3.3.1. Cleaning and Structuring

When constructing the function semantic framework, it is difficult to accurately get operator call frequencies with a rule-based traversal of raw scripts due to the diversity in user habits. Therefore, the code needs to be converted into a structured representation to design effective traversal algorithms for precise operator statistics. The Abstract Syntax Tree (AST) technology is introduced to represent the code in a structured form. The AST is a tree-like data structure that represents the syntactic structure of a program, clearly showing the hierarchical relationships between code components. Since geospatial scripts are based on languages like Java or Python, they can be effectively converted into structured JSON format with AST parsing, as illustrated in

Figure 4.

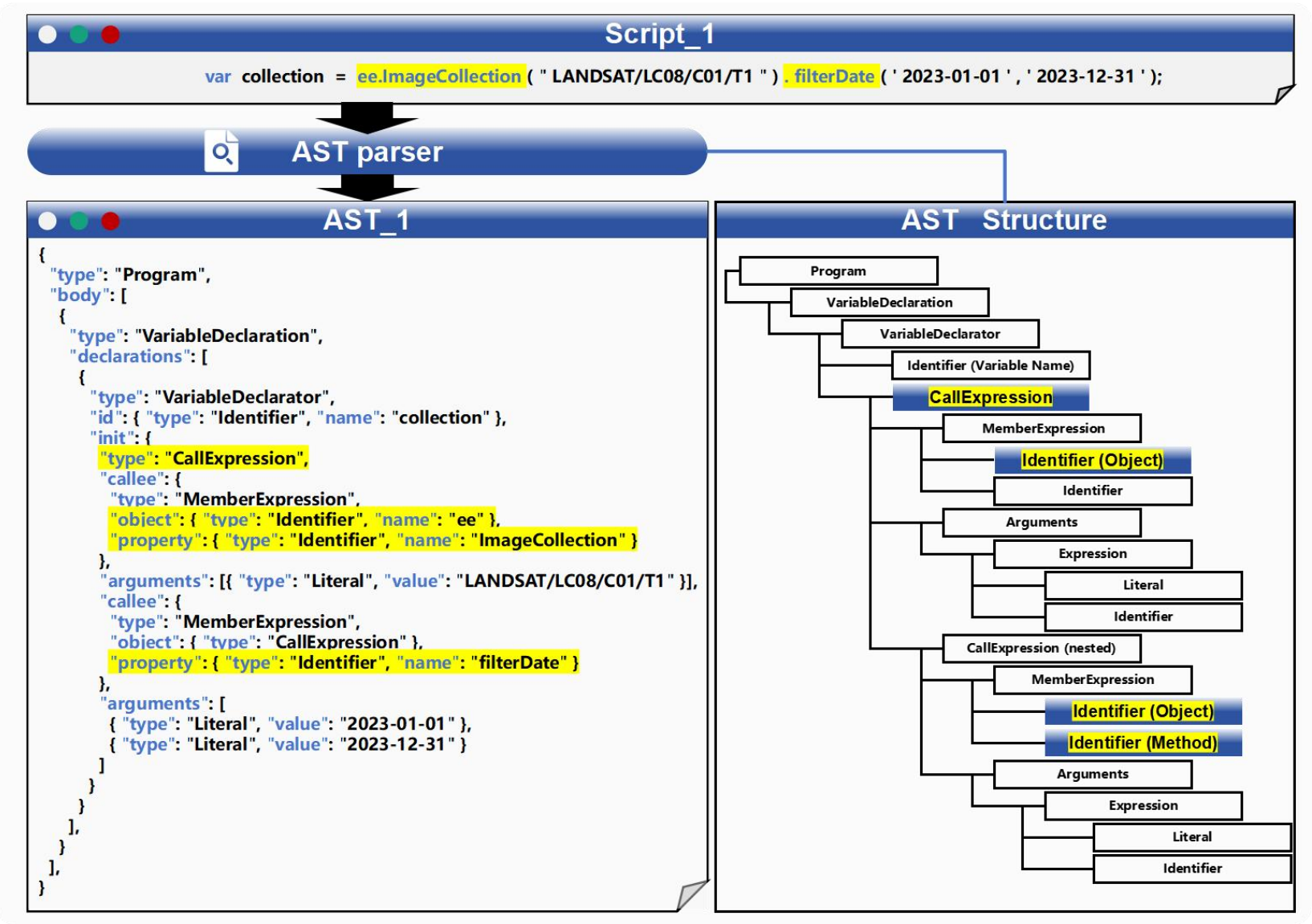

Fig.4 Schematic diagram of AST parsing process

In this structure, **Identifier** represents the operator name, **CallExpression** represents a function call, and **MemberExpression** represents the access of object properties or operators. The structured representation in ASTs not only clarifies the calling relationships between operators, but also facilitates the systematic counting of operator call frequencies through traversal algorithms, significantly enhancing the accuracy and efficiency of subsequent data processing and analysis.

### 3.3.2. Operator Relationship Frequency Statistics and Frequent Pattern Mining

Direct traversal methods show limitations in analyzing and identifying frequent operator combinations due to the uncertainty in combination lengths. In this study, a statistical method based on pairwise operator relationships is introduced. First, the pairing relationships and their occurrences between operators in each script are counted. Then, the same operator call relationships from different scripts are merged to form a comprehensive operator call frequency table. However, it is insufficient to merely count pairwise relationships and frequencies as the lengths of operator combinations are uncertain. Therefore, frequent pattern mining is necessary to identify high-frequency operator combinations, which provide essential data for mapping operator combinations to the function semantic framework and constructing the knowledge base.

The Apriori algorithm, a classic algorithm for frequent itemset and association rule mining[55], is used in this study. Its core principle is that any superset of a non-frequent itemset must also be non-frequent, which effectively reduces the search space. By applying the Apriori algorithm to the pairwise operator relationship frequency statistics, two-item sets can be extended to k-item sets. By using a support threshold, the meaningful frequent combinations are retained. The support of itemset X is calculated by:

$$Support(X) = \frac{\text{Frequency}(X)}{N} \#(12)$$

Where $\text{Frequency}(X)$ represents the number of occurrences of itemset X in the database, and N represents the total number of records in the database. Compared to traditional data mining methods, this approach only considers candidate itemsets that have already shown frequent occurrence as two-item combinations, then using them to build larger itemsets. This significantly reduces the search space, avoids ineffective calculations on low-frequency itemsets, and greatly optimizes the data processing workflow. It ensures the rational use of computational resources and the accuracy of analysis results on large-scale datasets. The pseudocode for operator relationship frequency statistics and frequent pattern mining is shown in Table 1.

Table.1 Pseudocode for Operator Frequency and Frequent Pattern Analysis

**Algorithm:Operator Frequency and Frequent Pattern Analysis**

**Input:** JSON files with ASTs for GEE scripts.

**Output:** List of frequent itemsets.

```
Function parse_AST(json_ast):
    log = []
    Traverse_AST(json_ast, log)
    Return calculate_frequencies(log)

Function traverse_AST(node, log, prev=None):
    If node.type == "CallExpression":
        current_call = extract_function_name(node)
        If prev: log.append((prev, current_call))
        prev = current_call
    For child in node.children: Traverse_AST(child, log, prev)
    If node.sequential_sibling:
        sibling_call = extract_function_name(node.sequential_sibling)
        If sibling_call: log.append((current_call, sibling_call))

Function extract_function_name(node):
    Return node.property.name if node.type == "MemberExpression" else node.name
```

**Algorithm:Operator Frequency and Frequent Pattern Analysis**

```
Function calculate_frequencies(log):
    call_pairs = {}
    For pair in log: call_pairs[pair] = call_pairs.get(pair, 0) + 1
    Return call_pairs

Function apriori(call_pairs, min_support):
    candidates = list(call_pairs.keys())
    freq_items = []
    While candidates:
        next_gen = []
        For itemset in candidates:
            support = sum(call_pairs.get(sub, 0) for sub in subsets(itemset)) / sum(call_pairs.values())
            If support >= min_support:
                freq_items.append(itemset)
                next_gen.extend(expand(itemset))
        candidates = list(set(next_gen))
    Return freq_items

Function subsets(itemset):
    Return [frozenset(x) for x in combinations(itemset, len(itemset)-1)]

Function expand(itemset):
    Return [itemset | {new_item} for new_item in all_items if new_item not in itemset]

Function main():
    json_ast = load_JSON_AST("path_to_ast.json")
    call_pairs = parse_AST(json_ast)
    freq_items = apriori(call_pairs, 0.05)
    Print freq_items

Call main()
```

As shown in Table 1, the **parse_AST** function firstly traverses the AST, records all pairwise operator call relationships and calculates their frequencies. Then, the Apriori algorithm is applied to filter frequent itemsets through a support threshold, expanding from two-item sets to k-item sets and generating new candidate itemsets. Core functions including **traverse_AST** for recursively traversing the AST and recording direct and sequentially adjacent operator pairs, **calculate_frequencies** for calculating frequencies, **subsets** for generating subsets, and **expand** for expanding candidate itemsets. Finally, the **main** function integrates these steps to complete frequent pattern mining and get the results.

### 3.4. Mapping Frequent Operator Combinations to the Function Semantic Framework

The statistical data of frequent operator combinations and the function semantic framework should be aligned so that the operator combinations can be accurately categorized into semantic functions. This issue involves mapping from statistical pattern recognition to semantic understanding. To achieve this, the LLM is employed for processing and understanding complex natural language texts. Furthermore, a Frequent Pattern Semantic Mapping (FPSM) technique that combines LLMs with voting mechanism is introduced for knowledge base construction, as illustrated in Figure 5.

The proposed Frequent Pattern Semantic Mapping (FPSM) technique is an innovative framework divided into four key steps: **Operator Sequence Orchestration and Distribution**, **Semantic Framework Indexing and Embedding Guidance**, **LLM-Guided Mapping Initialization**, **Existence Verification and Fuzzy Matching**.

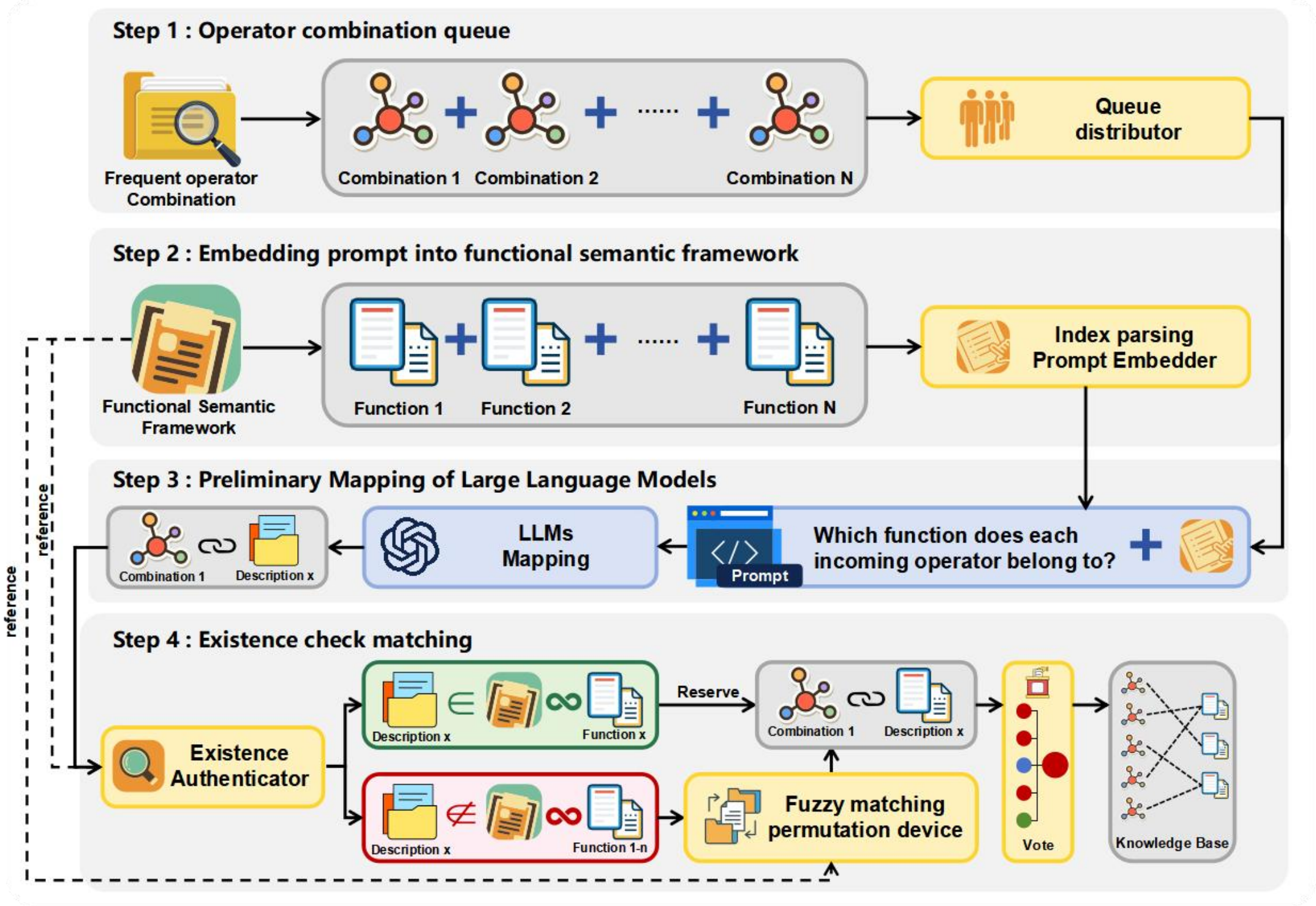

Fig.5 Frequent Pattern Semantic Mapping Diagram

In the **Operator Sequence Orchestration and Distribution**, high-frequency operator combinations are distilled from the set of frequent ones, which serve as the fundamental elements for subsequent mapping tasks. A sequence orchestration mechanism is designed to leverage the contextual memory and reasoning capabilities of LLMs, which employs a queuing distributor that organizes combinations by length, frequency and similarity. It ensures the mapping process starts with short and high-frequency combinations, gradually transitioning to more complex ones, effectively creating an implicit chain of thought that follows the processing logic of LLMs. Then in the **Function Semantic Framework Structuring and Prompt Construction**, the function semantic framework is converted into structured natural language, coupled with an indexing mechanism to turn the complex semantic framework into a set of prompt words for LLMs. This mechanism simplifies information expression and reduces the input token volume in LLM processing, thereby enhancing the efficiency and accuracy of function mapping. In the **LLM-Guided Mapping Initialization**, LLMs evaluate the input operator combinations and assign preliminary function descriptions within the function semantic framework, under the constraints of prompt words. However, the probabilistic nature of LLM generation may lead to unstable function descriptions. To address this, FPSM introduces an **Existence Verification and Fuzzy Matching Integration** mechanism. The existence verifier checks whether the function terms returned by the LLM match the terms within the semantic framework. For mismatches, fuzzy matching techniques are employed to select the closest term from the framework for mapping, ensuring standardized and consistent output. Finally, another prompt is designed and four advanced LLMs are selected for five consecutive inquiries. Using a voting mechanism through single-model or model combination schemes, the final mapping results are chosen, and ultimately completing the construction of the knowledge base.

## 4. Experimental Results

Based on the proposed method, an instance of Geo-SORB knowledge base is developed using geospatial scripts in GEE. To evaluate the construction result, $I_{\text{Geo-FuB}}$ metric is also designed to measure its semantic correctness and structural consistency.

### 4.1. Dataset and Experimental Setup

The dataset used in this study consists of 154,075 syntactically-checked GEE scripts. These scripts are written in the JavaScript programming language, spanning from September 2015 to September 2023. The size of individual scripts ranges from 1KB to 533KB, with a total dataset size of 0.799GB. The detailed characteristics of the dataset are shown in Table 2.

Table 2 : Characteristics of the GEE Script Dataset

| Attribute | Description |
|---|---|
| Programming Languages | JavaScript |
| Time Span | Sep.2015 - Sep.2023 |
| Number of Scripts | 154075 |
| Script Size Range | 1kb-533kb |
| Total Dataset Size | 0.799 GB |

The LLMs used in Geo-FuSE and Geo-FuM and their parameter settings are shown in Table 3.

Table 3 : Usage and Parameter Settings Table for Large Language Models

| Phase | Model Used | Parameter |
|---|---|---|
| Functional Semantic Framework Construction | GPT-4 | temperature=0.2<br>top-p=1 |
| Mapping | GPT-4<br>Llama 3-8B<br>ERNIE-4.0-8K<br>ERNIE-Speed-128K | |

The GPT-4 model is employed in the Geo-FuSE stage, while GPT-4, Llama 3-8B, ERNIE-4.0-8K, and ERNIE-Speed-128K were used for the operator combination and function framework mapping, as these models are leading in their respective fields at the time of writing. Specifically, in terms of parameter settings for the LLMs, the temperature was set to 0.2 and top-p is set to 1 to reduce the randomness of the output, making the models more likely to predict results with higher probabilities when generating text.

## 4.2. Semantic Framework Construction

### 4.2.1. Preliminary Quantitative and Qualitative Analysis

Based on 154,075 GEE geospatial scripts, the function semantic framework was constructed using GPT-4, where 520,335 functional statements are parsed. The quantitative analysis results of the script functional statements are illustrated in Figure 6.

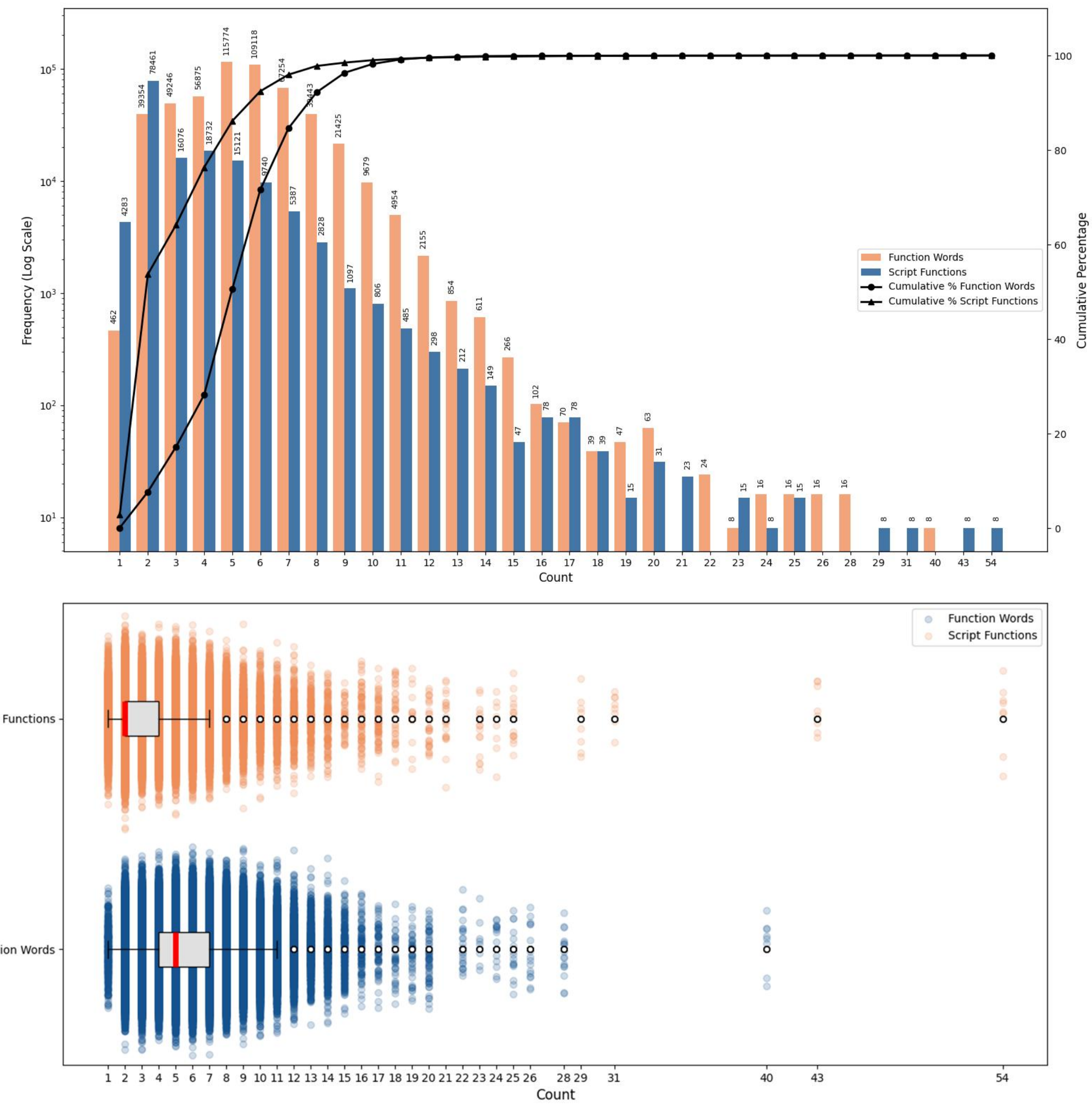

Fig.6 Quantitative Results Visualization of Functional Term Parsing

The number of functional statements parsed from individual scripts reveals the functional granularity distribution of the scripts. Analysis results show that the distribution of the number of functional statement within scripts is uneven, with more than 86.12% of scripts consisting of 1 to 5 functional statements, and 99.01% of scripts consisting of no more than 10 functional statements, reflecting users' preference for concise scripts. Quartile analysis (Q1=2, Median=2, Q3=4) further confirms this simplification trend but also shows a long-tail effect, with a few complex scripts containing up to 54 functional statements, illustrating the diversity and complexity of functional requirements.

Additionally, the number of words in functional statements reflects the linguistic conciseness of the LLMs when describing functions. 95.39% of functional descriptions use 9 or fewer words, with 49.10% using 1-5 words and 46.29% using 6-9 words, suggesting that concise descriptions effectively convey functions, which aids in standardization and consistency, and facilitate the semantic analysis and information extraction. However, 4.62% of descriptions exceed 10 words, with the longest reaching 40 words. This positively skewed distribution may lead to redundancy and inconsistency, increasing the hallucination phenomenon in LLMs. Overly descriptions lead to increased data dimensions, which in turn raises model complexity and computational cost. Therefore, adopting a standardized, concise description strategy can reduce redundancy and enhance the accuracy and consistency of information parsing. Word frequency statistics were performed on the parsed functional statements, a total of 8,232 different phrases were identified. The generated word cloud presents the proportional distribution of keywords in the functional statements of GEE scripts, as shown in Figure 7.

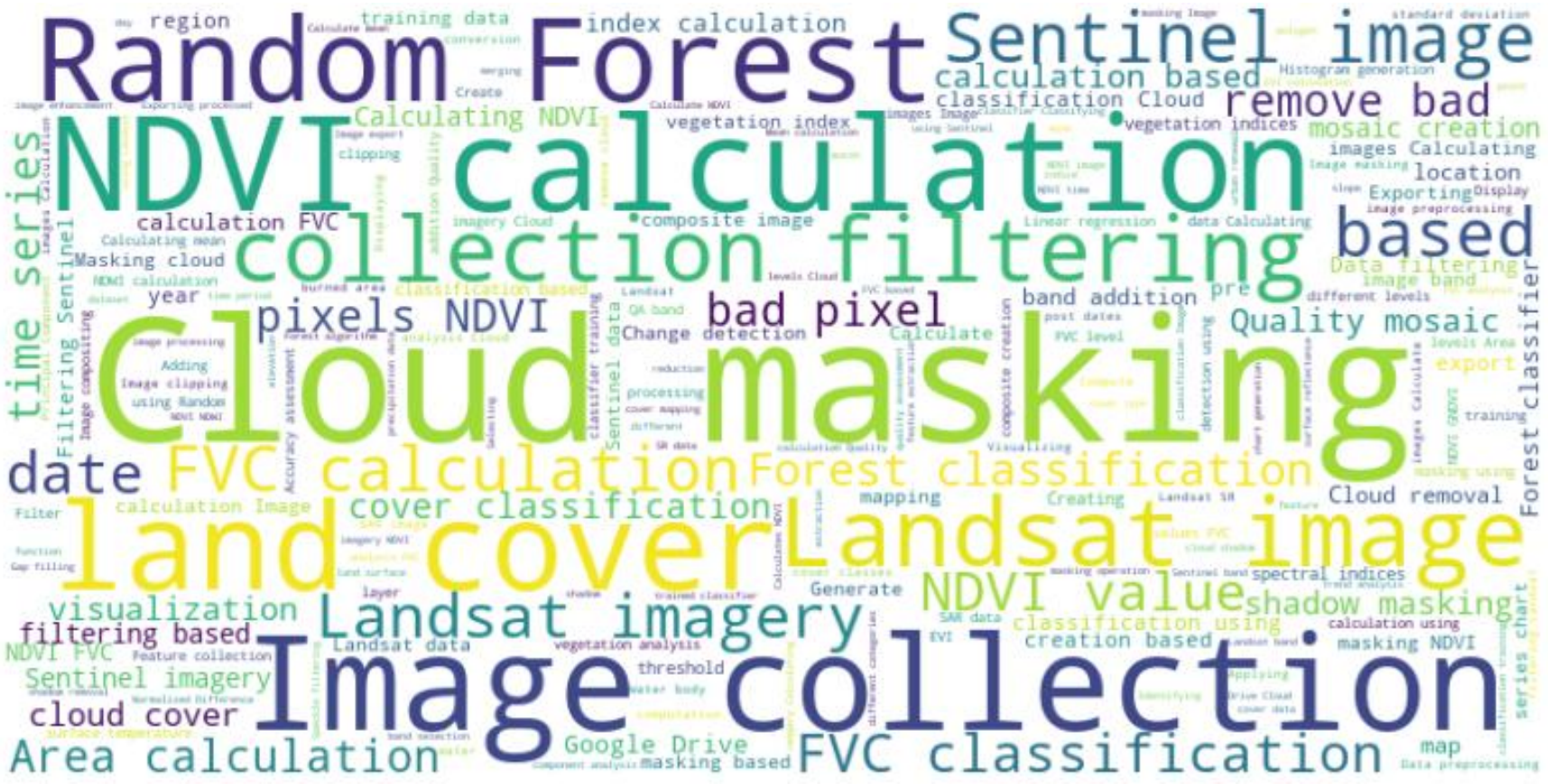

Fig.7 Functional word cloud

Furthermore, we conducted a quantitative analysis of high-frequency words, and the results are shown in Table 4. The statistics indicate that developers on the GEE platform primarily process data from Landsat, Sentinel-2, NDVI, FVC, and the bands of the images. They mainly focus on land cover and spatial area analysis. Therefore, the common operations include computation, masking, classification, detection, and filtering, which often involve integrated algorithms such as random forests. Additionally, many functions involve the composition of several basic operators. These results help identify key functions and assign higher attention and weight to related operations such as indicator analysis and land cover in subsequent expert annotations.

Table 4 : High frequency word statistics table

| Category | High-Frequency | Meaning | Frequency |
|---|---|---|---|

| Category | High-Frequency | Meaning | Frequency |
|---|---|---|---|
| Data Sets and Sources | Landsat | Analysis with Landsat imagery | 5398 |
| | Sentinel-2 | Analysis with Sentinel-2 imagery | 2926 |
| | NDVI | NDVI calculation and analysis | 7810 |
| | FVC | FVC calculation and analysis | 3593 |
| Analysis Scope | Band | Band processing and index calculations | 1890 |
| | Cover | Land cover and area calculations | 3734 |
| | Area | Area analysis for regions | 1327 |
| Data Processing and Analysis Tasks | Calculation | Numerical calculations and transformations | 1419 |
| | Masking | Data cleaning, e.g., cloud removal | 6730 |
| | Classification | Land cover classification or object recognition | 4854 |
| | Detection | Feature extraction and change detection | 1319 |
| | Filtering | Data filtering and area selection | 3214 |
| Methods and Tools | Random Forest | Using Random Forest for classification | 1904 |
| Frequency of Terms and Keywords | for | Used in loops and conditions | 18412 |
| | and | Used in logical connections | 17576 |
| | to | Indicating data flow or steps | 5752 |
| | on | Describing position or conditions | 5715 |
| | of | Connecting processing and analysis tasks | 5065 |

#### 4.2.2. Word Vector Construction and Clustering

The parameters used for word vector construction and clustering are shown in Table 5:

Table 5 : High frequency word statistics table

| Parameter | Value/Default |
|---|---|
| perplexity | 30 |
| n_iter | 3000 |
| random_state | 10 |
| scaler | StandardScaler() |

In the t-SNE algorithm, parameter perplexity is used to balance the local and global data structure, with a commonly used value of 30. Parameter **n_iter** sets the number of iterations for optimizing the low-dimensional embedding, with higher iteration counts (such as 3000) helping to improve convergence. Parameter **random_state** provides a seed for the random number generator in both t-SNE and GMM algorithms, ensuring the reproducibility of results, which allows for consistent output and easier comparison of results across different runs. Parameter **scaler**, specifically **StandardScaler**, is used to standardize the data, ensuring that each feature has equal influence in the analysis. The standardized data has a mean of 0 and a variance of 1, which is crucial for t-SNE and GMM as these algorithms are sensitive to the scale of input features.

The TF-IDF values for 520,335 functional statements were calculated and t-SNE is performed for dimensionality reduction to generate low-dimensional vector points. By computing the BIC values for GMM clustering with the number of word vector clusters ranging from 1 to 500, the optimal number of clusters is identified as 42, as shown in Figure 8. Since the “elbow point” occurs at 42, where the BIC value is minimized, the inter-cluster distance is maximized, the intra-cluster distance is small, and the computational redundancy is reduced. Therefore, we chose to cluster the word vectors into 42 clusters.

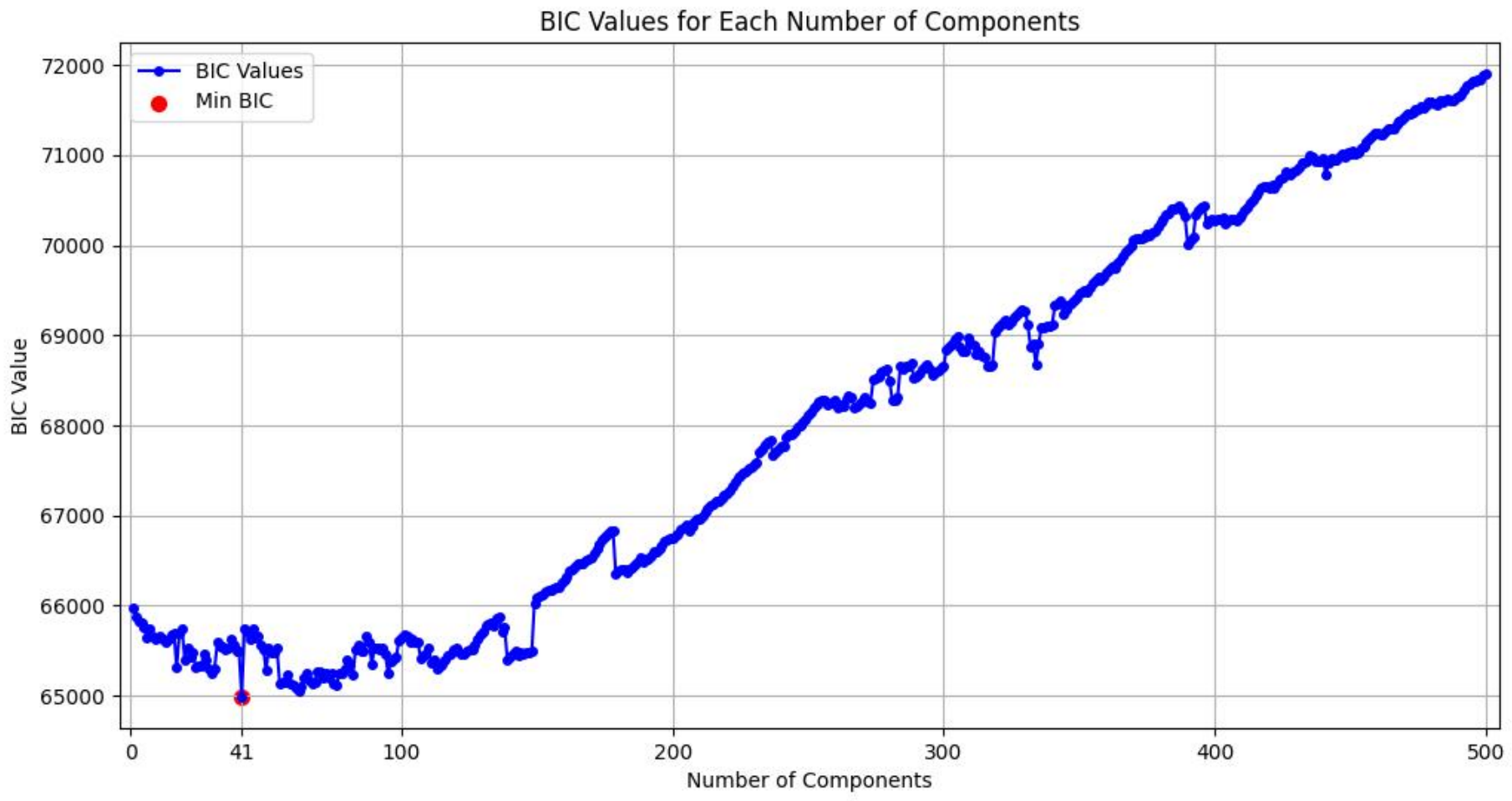

Fig.8 BIC Values of Gaussian Mixture Model

The visualization for the clustering of the 520,335 word vectors in the low-dimensional space is shown in Figure 9. The result indicates the clear boundaries and internal structures of different clusters. Different colors and shapes are used to distinguish the clusters, highlighting the effectiveness of GMM in handling high-dimensional data. The grouping characteristics of each functional statement can be intuitively observed through this visualization.

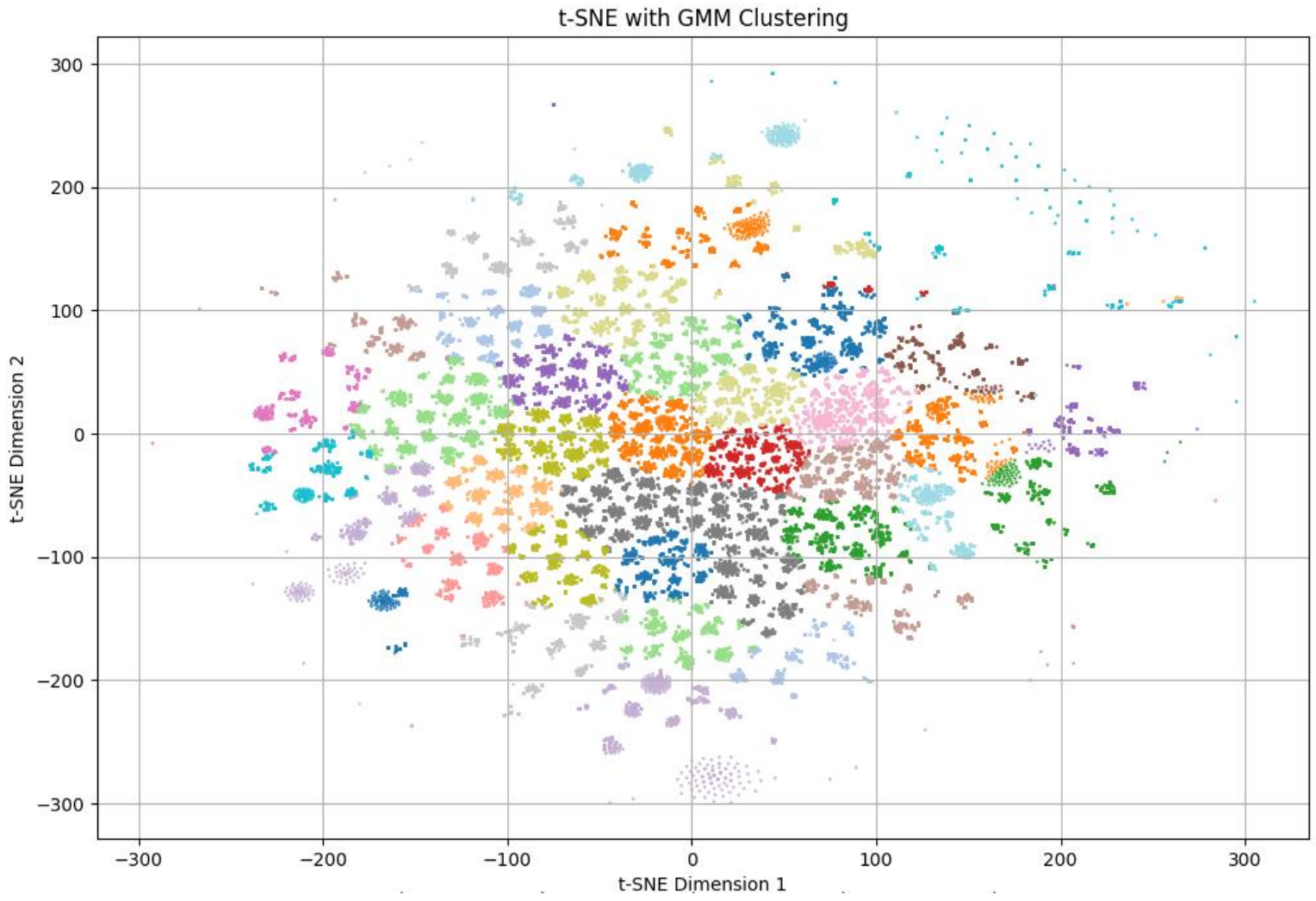

Fig.9 Visualization of Word Vectors Clustering in 42 Categories

In the semantic assignment stage with expert experience, certain clusters shew a clear data source inclination, such as "clipping Landsat" and "clipping Sentinel-2". These clusters indicated specific data sources for functional operations. However, the operator combinations do not contain data source information and are not specifically designed for a particular data source. Therefore, although these functions are reasonable in semantics, they do not align with the task goal of "assigning functional semantics to operator combinations" presented in this study, while the objective is to accurately map function descriptions to the operator level without involving specific data sources. Additionally, we found that some clusters had similar semantics and could be merged. For example, "data preprocessing" and "data cleaning" have a high degree of semantic overlap. After careful analysis and revision, a three-level semantic framework was constructed, including 3 categories, 8 subcategories, and 21 specific functional terms, as shown in Table 6.

Table 6 : Functional Semantic Framework

| Id | Category | Subcategory | Specific Function |
|---|---|---|---|
| 1 | Data Preprocessing | Data Preparation | Data Import |
| 2 | | | Data Filtering |
| 3 | | Image Processing | Image Mosaicking and Blend |
| 4 | | | Clip Images |
| 5 | | | Image Masking |

| Id | Category | Subcategory | Specific Function |
|---|---|---|---|
| 6 | Core Spatiotemporal Analysis | Numerical Calculation | Image Correction |
| 7 | | | Image Band Manipulation |
| 8 | | | Normalization of Results |
| 9 | | | Area Calculation |
| 10 | | | Image Index Calculation |
| 11 | | | Feature Calculation |
| 12 | | | Threshold Application |
| 13 | | Image Processing | Buffer Analysis |
| 14 | | | Analysis of Unicom Zone |
| 15 | | Machine Learning | Cluster Analysis |
| 16 | | | Classification and Machine Learning |
| 17 | Data Post-Processing | Data Processing | Smoothing Results |
| 18 | | | Resampling Results |
| 19 | | | Data Conversion |
| 20 | | Data Visualization | Chart Visualization |
| 21 | | Data Export | Export Data |

### 4.3. Operator Combination Results

#### 4.3.1. Operator Frequency Statistics Results

The GEE platform's operator library includes 1,395 operators. Based on the frequency statistics on the dataset, 3,317 different operator pairs are identified, representing 3,317 unique call patterns, covering 322 operators. We defined 190 calling operators and 306 called operators according to the call sequence. The visualization of operator call relationships is shown in Figure 10, in which nodes represent operators, and edges represent call relationships. Larger nodes indicate operators with more call relationships, and thicker edges indicate higher call frequency.

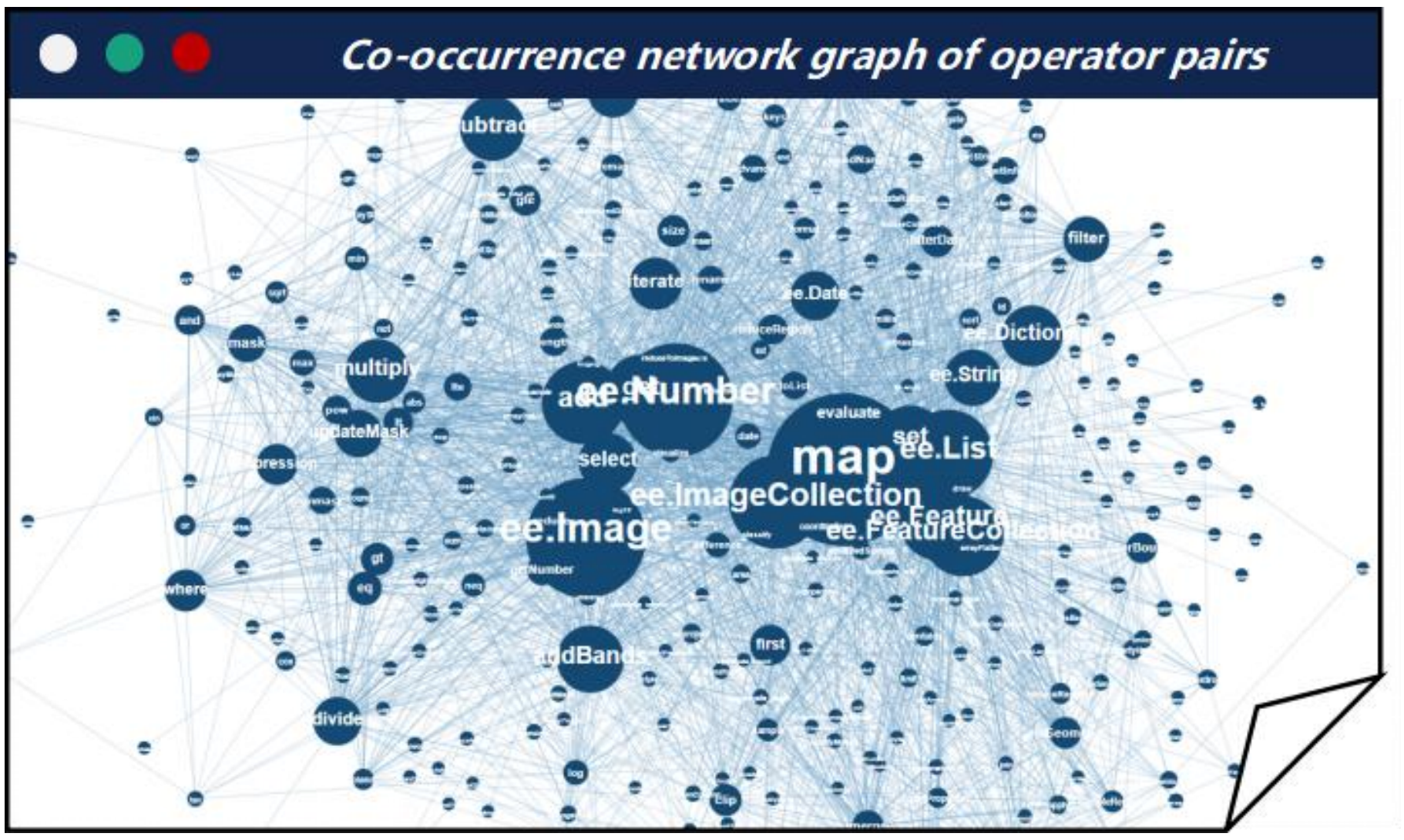

Fig.10 Diagram of Operator Invocation

Quantitative analysis of high-frequency operator pairs reveals the primary needs and operational patterns of GEE users. By analyzing high-frequency operator pairs, we obtained the top 20 operator pairs by frequency, as shown in Table 7.

Table 7 : High-frequency operator pair statistics table

| Id | Caller | Callee | Frequency |
|---|---|---|---|
| 1 | expression | select | 520690 |
| 2 | ee.Image | get | 469870 |
| 3 | ee.Number | get | 372491 |
| 4 | addBands | select | 325159 |
| 5 | map | ee.Image | 215731 |
| 6 | map | ee.Number | 213305 |
| 7 | map | ee.Number | 201195 |
| 8 | updateMask | eq | 193692 |
| 9 | set | get | 172411 |
| 10 | map | select | 163319 |
| 11 | ee.Array | ee.Dictionary | 156305 |
| 12 | addBands | reduceNeighborhood | 154139 |
| 13 | toList | size | 151029 |
| 14 | ee.Date | get | 142028 |
| 15 | addBands | rename | 137412 |

| Id | Caller | Callee | Frequency |
|---|---|---|---|
| 16 | where | gt | 131434 |
| 17 | ee.List | get | 121122 |
| 18 | updateMask | gt | 115347 |
| 19 | multiply | subtract | 106367 |
| 20 | addBands | addBands | 105120 |

The analysis results indicate that these high-frequency operator pairs primarily involve the categories of data selection and extraction, data computation and transformation, and conditional filtering and screening operations. For example, operator pairs “expression-select” and “ee.Image-get” play crucial roles in data selection, extraction and preprocessing, while “map-ee.Image” and “addBands-reduceNeighborhood” are widely used in image analysis and data transformation. Additionally, operator pairs “updateMask-eq” and “where-gt” are used for conditional filtering and screening of data. By conducting a slice analysis based on call frequency, a strong correlation between high-frequency and low-frequency operators can be observed, making them suitable for the mining of frequent operator combinations, as illustrated in Figure 11.

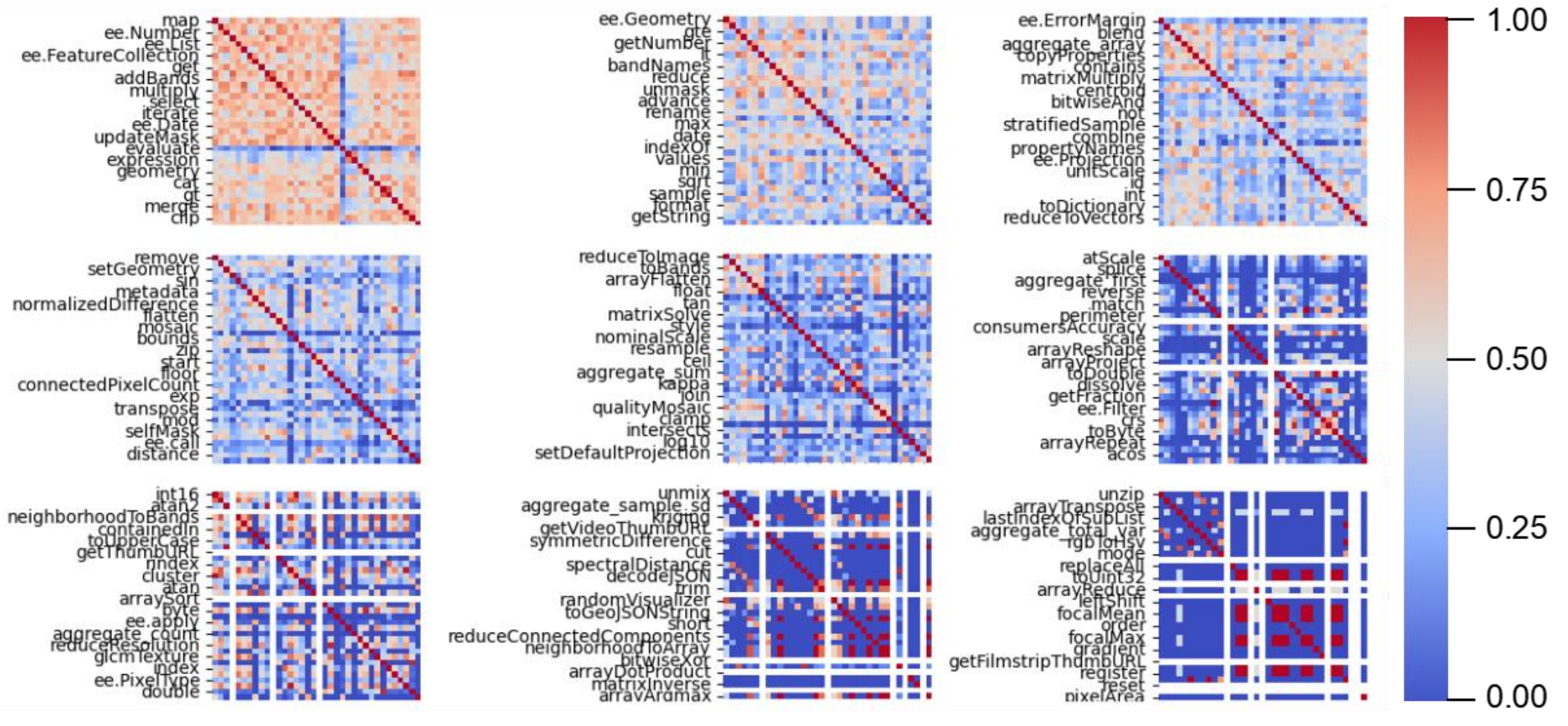

Fig.11 Correlation heat map between operators

#### 4.3.2. Frequent Operator Combination Mining

By applying the Apriori algorithm to the 3,317 unique call patterns and setting the minimum support   to 0.05, only itemsets with a support of at least 5% were considered frequent itemsets. This process identified 4,350 frequent itemsets, referred to as frequent operator combinations. The distribution of the number of operators within these combinations is shown in Table 8.

Table 8 : Distribution of Operators in Frequent Itemsets

| Number of Operators | Number of Frequent Itemsets |
|---|---|
| 1 | 51 |
| 2 | 322 |
| 3 | 850 |
| 4 | 1202 |
| 5 | 1038 |
| 6 | 587 |
| 7 | 227 |
| 8 | 61 |
| 9 | 11 |
| 10 | 1 |

The correlation between the size of itemsets and their frequency follows a distinct trend: a mere 51 frequent itemsets consist of single operators, suggesting that standalone operators rarely meet the support threshold for frequent itemsets. This implies that most application scenarios require combinations of multiple operators. For small itemsets with 2 to 4 operators, the frequency is notably higher, with a pronounced peak at itemsets with 4 operators, reaching the number 1,202. This underscores the ubiquity and significance of such combinations across a variety of application scenarios. Itemsets of moderate size with 5 to 6 operators, are found in quantities of 1,038 and 587, respectively. Although there is a decline compared to smaller itemsets, they still demonstrate considerable usability in handling tasks of moderate complexity. The count of large itemsets, which include 7 to 10 operators, experiences a marked drop, with combinations of 9 and 10 operators being exceedingly rare, at 11 and 1 itemsets. This indicates that exceedingly intricate operator combinations are not commonly encountered in most application scenarios. This trend reflects the long-tail distribution which is often observed in the realm of frequent itemset mining.

### 4.4. Mapping Results

Four LLMs, including GPT-4, Llama 3-8B, ERNIE-4.0-8K, and ERNIE-Speed-128K, are used for mapping frequent operator combinations into the function semantic framework. The experimental setup is as follows: **(1)** The operator combinations were sorted based on combination length and call frequency before being input into each model. The three-level semantic framework was embedded into the prompt. **(2)** Five consecutive mappings performed independently with each of the four LLMs, and the results of these five mappings were voted on to determine each LLM's final functional label output. **(3)** The outputs of each model were further integrated, including pairwise combinations (10 mapping results), three-model combinations (15 mapping results), and all four models combined (20 mapping results), with a comprehensive vote taken to obtain the final integrated result. We generate a total of 15 model combination results through the experiments, which will be further evaluated.

To assess the structural consistency and semantic accuracy of the model mapping results, we used expert annotations as the ground truth for verification, labeling 150 common operator combinations. A comprehensive sampling strategy was employed to ensure the representativeness of samples. Initially, random sampling was conducted based on the proportion of operator combination sequence lengths. To examine the models' ability to recognize small differences, some similar combinations, differing by only one operator, were specifically selected. Additionally, given the differences in operator occurrence frequency, some combinations of low-frequency operators are particularly chosen. Each sample was labeled by three experts for its functional category, samples with ambiguous functions were excluded based on the annotation results to ensure the accuracy of the annotations.

A scientific metrics system for evaluating verification results was designed in this study, which is divided into structural consistency metrics (weighted 40%) and semantic correctness metrics (weighted 60%). The evaluation is based on three-level labels, where a Boolean value $T_i$ is used to represent the correctness of the $i\text{-}th$ level label. The semantic correctness metric $I_{semantic}$ is calculated independently of the structural consistency metric $I_{structure}$. The semantic correctness metrics $I_{semantic}$ is calculated as follows:

- First, the correctness of the third-level label is verified: if the third-level label is completely correct, the semantic correctness metric scores full points.
- If the third-level label is incorrect, regardless of whether the first and second-level labels

are consistent, the semantic correctness metric score is determined by the semantic similarity calculation algorithm, i.e., the semantic similarity multiplied by the full score of the semantic correctness metric. The semantic similarity is calculated as follows, where $S$ is the semantic similarity, ranging from 0 to 1:

$$I_{\text{semantic}} = \begin{cases} 100, & \text{if } T_3 = \text{True} \\ S \times 100, & \text{if } T_3 = \text{False} \end{cases} \#(13)$$

Similarly, the structural consistency metric $I_{structure}$ is calculated as follows:

- If the third-level label is completely correct, the structural consistency metric scores full points.
- If the third-level label is incorrect, the correctness of the second-level label is checked:
  - If the second-level label is correct, the score is $(\text{Full Score} \times 80\%)/n$, where $n$ is the number of third-level indicators under this correct second-level indicator.
  - If the second-level label is incorrect, the first-level label is checked. If the first-level label is correct, the score is $(\text{Full Score} \times 20\%)/m$, where $m$ is the number of second-level indicators under this correct first-level indicator.

$$I_{\text{structure}} = \begin{cases} 100, & \text{if} T_3 = \text{True} \\ \frac{80}{n}, & \text{if} T_3 = \text{False} \wedge T_2 = \text{True} \\ \frac{20}{m}, & \text{if} T_3 = \text{False} \wedge T_2 = \text{False} \wedge T_1 = \text{True} \\ 0, & \text{if} T_3 = \text{False} \wedge T_2 = \text{False} \wedge T_1 = \text{False} \end{cases} \#(14)$$

The final evaluation metric $I_{\text{Geo-SORB}}$ is the weighted sum of the semantic correctness metric score and the structural consistency metric score:

$$I_{\text{Geo-FuB}} = 0.6 \times I_{\text{semantic}} + 0.4 \times I_{\text{structure}}$$

This metric design ensures a comprehensive evaluation of the model mapping results, combining multi-level structural and semantic analysis, in line with data science evaluation standards. The final evaluation results are shown in Table 9.

Table 9 : Evaluate Results

| **Model Combination** | $I_{\text{structure}}$ | $I_{\text{semantic}}$ | $I_{\text{Geo-SORB}}$ |
|---|---|---|---|
| GPT-4 | **85.297** | **83.231** | **84.057** |
| Llama 3-8B | 82.654 | 82.094 | 82.318 |

| Model Combination | $I_{structure}$ | $I_{semantic}$ | $I_{Geo\text{-}SORB}$ |
|---|---|---|---|
| ERNIE-4.0-8K | 84.135 | 82.489 | 83.147 |
| ERNIE-Speed-128K | 80.328 | 80.187 | 80.243 |
| GPT-4 + Llama 3-8B | 87.472 | 83.721 | 85.221 |
| GPT-4 + ERNIE-4.0-8K | **88.297** | **84.378** | **85.946** |
| GPT-4 + ERNIE-Speed-128K | 86.519 | 83.129 | 84.485 |
| Llama 3-8B + ERNIE-4.0-8K | 84.835 | 83.504 | 84.036 |
| Llama 3-8B + ERNIE-Speed-128K | 82.479 | 81.768 | 82.052 |
| ERNIE-4.0-8K + ERNIE-Speed-128K | 81.326 | 80.538 | 80.853 |
| GPT-4 + Llama 3-8B + ERNIE-4.0-8K | **90.517** | **85.724** | **87.641** |
| GPT-4 + Llama 3-8B + ERNIE-Speed-128K | 89.218 | 85.182 | 86.796 |
| GPT-4 + ERNIE-4.0-8K + ERNIE-Speed-128K | 89.764 | 85.348 | 87.114 |
| Llama 3-8B + ERNIE-4.0-8K + ERNIE-Speed-128K | 86.047 | 83.214 | 84.347 |
| **GPT-4 + Llama 3-8B + ERNIE-4.0-8K + ERNIE-Speed-128K** | **92.034** | **86.792** | **88.889** |

Evaluation results for representative LLMs, which balance performance and speed in various fields, indicate that GPT-4 excels in single-model usage, showcasing outstanding natural language processing capabilities. Although ERNIE-Speed-128K scores lower when used alone, it still plays a positive role when combined with other models. Overall, model combinations tend to significantly improve performance as they integrate the strengths of different models, providing more comprehensive and stable evaluation results. When using all four LLMs together, the best performance across various metrics is achieved, with a structural consistency metric score of 92.034, a semantic correctness metric score of 86.792, and a final evaluation metric score of 88.889. These results show the scientific rigor and effectiveness of the proposed construction method.

## 5. Discussions

### 5.1. Application Scenario of the Geo-FuB Knowledge Base

The construction method proposed in this paper can be applied to build knowledge bases for various geospatial platforms, and facilitate automated modeling tasks using LLMs. To adapt such knowledge bases for specialized geospatial code generation tasks, two main LLM-based paradigms are raised: (1) **Chain-of-Thought Retrieval-Augmented Generation:** It is suitable when the LLM has coding capabilities and geospatial knowledge but requires guidance. In this case, Geo-FuB can be used as an external knowledge base; (2) **LLM Fine-Tuning:** It is appropriate when the LLM lacks sufficient knowledge. In this sense, Geo-FuB can be used as a pre-training dataset for fine-tuning. This section focuses on discussing these two cutting-edge research paradigms and encourages further exploration and expansion in the downstream application of the proposed Geo-Fub Knowledge in this study, as illustrated in Figure 12.

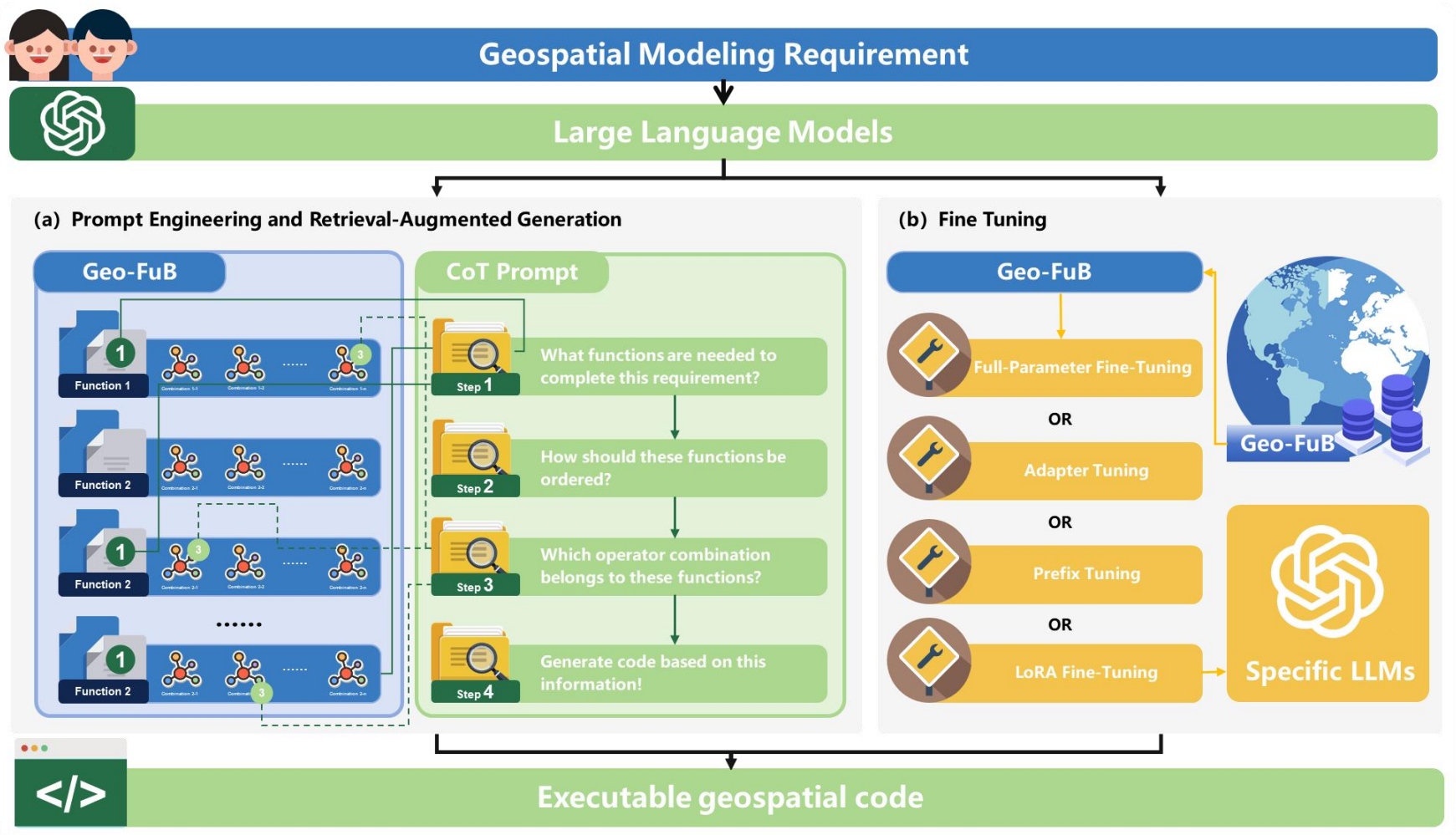

Fig.12 Using Geo-FuB for automatic geospatial code generation tasks in two LLM-based paradigms

#### 5.1.1. Chain-of-Thought Retrieval-Augmented Generation

In the geospatial modeling domain, applying LLMs directly to generate geospatial code often poses challenges because of the absence of highly specialized domain knowledge, the ambiguity of user requirements, and the potential misalignment between the model's comprehension and the task's unique demands. Therefore, it is essential to craft prompt engineering that employs a Chain-of-Thought (CoT), followed by the Retrieval-Augmented Generation (RAG) to access the external Geo-FuB knowledge base, thereby optimizing this knowledge-intensive geospatial code generation. The CoT method simulates the thought process of human teaching by setting a series of highly relevant prompt words, provides a heuristic thinking environment for LLMs, and guides them to progressively approach the correct answer. RAG enhances the model's retrieval capabilities by integrating a pre-constructed knowledge base, ensuring that the model accesses accurate and relevant information. The Geo-FuB knowledge base in this study is structured in JSON format with an ID and indexing mechanism, enabling the LLM to effectively index functional labels. Specifically, when faced with a user modeling demand, the task can be solved gradually through layered questions: (1) Ask the LLM which functions are involved; (2) Inquire about the order in which these functions should be combined; (3) Identify which specific operator combinations from these functions are needed for code generation. After responding step by step, the required geospatial code can be generated to realize the geospatial modeling objectives.

#### 5.1.2. Fine-Tuning Large Language Models

By incorporating domain-specific knowledge from the knowledge base through fine-tuning approaches, general LLMs can retain their general language processing capabilities while acquiring the specialized knowledge required for specific tasks. The fine-tuning methods mainly include Full-Parameter Fine-Tuning, Adapter Tuning, Prefix Tuning, and LoRA (Low-Rank Adaptation) Fine-Tuning. **Full-Parameter Fine-Tuning** involves retraining all the model parameters to ensure optimal performance in specific tasks, and requires substantial computational resources. **Adapter Tuning, Prefix Tuning and LoRA Fine-Tuning** reduce parameters to be retrained and computational resources required by inserting adapter modules into the model, adjusting the prefix part of the model input or adjusting the model's weight matrices through low-rank approximation, thus increasing the fine-tuning efficiency.

Compared to RAG technology, the process of fine-tuning a model demands a more substantial investment of computational resources. However, this method yields a higher degree of domain adaptability, which leads to improved accuracy and efficiency when tackling domain-specific tasks. Fine-tuning is particularly suitable for scenarios that call for profound domain expertise. Therefore, by fine-tuning LLMs with the Geo-FuB knowledge base, these LLMs can be specialized to precisely grasp the intricacies of terminology, logic, and operations of the geospatial modeling domain , thereby enhancing their proficiency in geospatial code genenration tasks.

### 5.2. Limitations

This study proposes an effective method for constructing the knowledge base that facilitates LLMs in automated geospatial code generation for the first time. By mining the semantic annotations and operator combination statistics from geospatial scripts, this method effectively guides LLMs to achieve better performance in the geospatial domain. However, the current study has several limitations. For example, the study primarily focuses on operator combinations without including parameter passing and dataset information as statistical indicators in the knowledge base. It only considers the existence of operators without addressing the impact of their connection order. To further improve the model's performance and applicability, future research should expand the scope of the knowledge base, thoroughly mining and fully utilizing the information from geospatial scripts. Additionally, the proposed function semantic framework could be further integrated with existing geospatial standard frameworks, such as the OGC (Open Geospatial Consortium) geoprocessing ontology, to enhance its interpretability of the functional division.

## 6. Conclusion and Future Work

This paper proposes the Geo-FuB framework for constructing an operator-function knowledge base for LLMs in the geospatial code generation task. The framework includes three core components: Function Semantic Framework Construction (Geo-FuSE), Frequent Operator Combination Statistics (Geo-FuST), and Combination and Semantic Framework Mapping (Geo-FuM), which are suitable for knowledge base construction of various geospatial analysis platforms. Based on 154,075 geospatial scripts in Google Earth Engine (GEE) , a Geo-FuB instance is constructed that includes a semantic framework with 3 primary categories, 8 secondary categories, and 21 tertiary categories, as well as 4,350 sets of frequent operator combinations and their complete knowledge base mapping results. This knowledge base instance is now open source on https://github.com/whuhsy/Geo-FuB and ready-for-use in code generation and other downstream tasks. We proposed a comprehensive evaluation metric that considers both structural integrity and semantic accuracy. The evaluation results show an overall accuracy of 88.89%, with structural accuracy reaching 92.03% and semantic accuracy at 86.79%. Through various application scenarios, this paper highlights how Geo-FuB optimizes the automatic generation of geospatial modeling scripts in two cutting-edge research paradigms: Retrieval-Augmented Generation (RAG) and fine-tuning, which encourages further exploration and expansion of the knowledge base in downstream applications.

Future research may consider the sequence of operators in their combinations and establish an external knowledge base using graph structures. The potential and feasibility of using graph structures in LLMs for geospatial code generation tasks will be explored. Additionally, we plan to establish related geospatial code datasets and use the knowledge base proposed in this study to fine-tune open-source general LLMs for geospatial code generation.

## Acknowledgements

This work was supported by the National Natural Science Foundation of China under Grant No. 41930107, awarded to Huayi Wu.